\documentstyle[aps,epsf]{revtex}

\preprint{NUC-MINN-01/9-T}

\newcommand{\e}[2]{\epsilon_{#1}^{#2}}
\newcommand{\dinot}{\partial\!\!\!/}
\newcommand{\Ahatnot}{\hat{A}\!\!\!/}
\newcommand{\vpsl}{\vec{p}\!\!\!/}
\newcommand{\be}{\begin{equation}}
\newcommand{\ee}{\end{equation}}
\newcommand{\ba}{\begin{eqnarray}}
\newcommand{\ea}{\end{eqnarray}}

\preprint{NUC-MINN-01/9-T}

\begin{document}
\draft

\title{Masses of the pseudo-Nambu-Goldstone bosons
in two flavor color superconducting phase}

\author{V.A.~Miransky$^{*}$}

\address{Department of Applied Mathematics, University of Western 
Ontario, London, Ontario N6A 5B7, Canada}

\author{I.A.~Shovkovy$^{*}$}

\address{School of Physics and Astronomy, University of Minnesota,
Minneapolis, MN 55455, USA}

\author{L.C.R.~Wijewardhana}
\address{Physics Department, University of Cincinnati,
Cincinnati, Ohio 45221-0011, USA}

\date{\today}
\maketitle

\begin{abstract}

The masses of the pseudo-Nambu-Goldstone bosons in the color
superconducting phase of dense QCD with two light flavors are estimated by
making use of the Cornwall-Jackiw-Tomboulis effective action.
Parametrically, the masses of the doublet and antidoublet bosons are
suppressed by a power of the coupling constant as compared to the value of
the superconducting gap. This is qualitatively different from the mass
expression for the singlet pseudo-Nambu-Goldstone boson, resulting from
non-perturbative effects. It is argued that the (anti-) doublet
pseudo-Nambu-Goldstone bosons form colorless [with respect to the unbroken
$SU(2)_{c}$] charmonium-like bound states. The corresponding binding
energy is also estimated.

\end{abstract}
\pacs{11.15.Ex, 12.38.Aw, 12.38.-t, 26.60.+c}


\section{Introduction}
\label{intro}

Although there is no reliable observational signature yet, from
theoretical point of view, it is quite reasonable to assume that the cores
of compact stars are made of color superconducting quark matter
\cite{Alford:2000sx}. If we take this assumption seriously, it becomes
quite important to study the properties of the possible color
superconducting phases in full detail (for an up to date review on color
superconductivity see Refs.~\cite{Rajagopal:2000wf,Alford:2001dt}). In
this paper, we continue our study \cite{Miransky:2000gg,Miransky:2000jt}
of the pseudo-NG bosons, related to the approximate axial color symmetry,
in the $S2C$ phase of cold dense QCD.

Let us recall that the axial color transformation is not a symmetry of the
QCD action. However, its explicit breaking is a weak effect at
sufficiently high quark densities where the coupling constant
$\alpha_{s}(\mu)$ is small ($\mu$ is a chemical potential). This was our
main argument in Refs.~\cite{Miransky:2000gg,Miransky:2000jt} suggesting
the existence of five (rather than one \cite{Beane:2000ms}) light
pseudo-NG bosons in the $S2C$ phase of cold dense QCD. No mass estimates for
these pseudo-NG bosons were given in
Refs.~\cite{Miransky:2000gg,Miransky:2000jt}. In this paper, we fill in
the gap by developing the formalism and calculating the masses.

This paper is organized as follows. In the next section, we briefly
introduce our model and notation.  In Sec.~\ref{formalism}, we describe
our method, based on the Cornwall-Jackiw-Tomboulis (CJT) effective action,
for calculating mass estimates of the pseudo-NG bosons. Then, in
Sec.~\ref{vac-diagram}, the leading order diagram is approximately
calculated, using analytical methods. The fate of the colored pseudo-NG
bosons in the doublet and antidoublet channels is discussed in
Sec.~\ref{2-confinement}.  In Sec.~\ref{conclusion}, we give our
conclusions.  In Appendices~\ref{pol-tensor} and \ref{integrals} we
present the general expression for the gluon polarization tensor and the
calculation of the integrals that appear in its definition.
In Appendix C the problem of the gauge invariance in the loop
expansion of the CJT effective action is discussed. 

\section{The Model and Notation} 
\label{Model}

Here we consider cold dense QCD with two light quark flavors ($u$ and $d$)
in the fundamental representation of the $SU(3)_{c}$ color gauge group. In
order to keep the analytical calculation under control, we assume that the
chemical potential $\mu$ is much larger than $\Lambda_{QCD}$. Of course,
when we talk about the compact stars, this is a far stretched assumption.
Therefore, while extending our analytical results to the realistic
densities existing, for example, at the cores of compact stars, one should
be very careful. While most of the qualitative results may survive without
being affected, most of the quantitative estimates would probably be valid
only up to an order of magnitude. From the viewpoint of a theorist, it is
still most interesting to study the predictions of the microscopic theory,
{\em i.e.} QCD in the problem at hand. The price for such a luxury is the
necessity to work at asymptotically large densities.

Instead of working with the standard four component Dirac spinors, in our
analysis below, it is convenient to introduce the following eight
component Majorana spinors: 
\begin{equation}
\Psi=\frac{1}{\sqrt{2}}\left(\begin{array}{c}\psi_{D} \\
       \psi^{C}_{D}
   \end{array}\right), \quad
\psi^{C}_{D}=C\bar{\psi}_{D}^{T},
\label{Psi}
\end{equation}
where $\psi_{D}$ is a Dirac spinor and $C$ is a charge conjugation 
matrix, defined by $C^{-1} \gamma_{\mu}C =-\gamma_{\mu}^{T}$ and 
$C=-C^{T}$. In this notation, the inverse fermion propagator
in the color superconducting phase reads
\cite{Hong:2000fh,Schafer:1999jg,Pisarski:2000bf,Hsu:2000mp,Brown:2000aq}
\begin{eqnarray}
\left(G(p)\right)^{-1}&=&-i\left(\begin{array}{cc}
(p_0+\mu)\gamma^{0}+\vpsl & \Delta\\
\tilde{\Delta} & (p_0-\mu)\gamma^{0}+
\vpsl  \end{array}\right) \nonumber \\
&=&-i\left(\begin{array}{cc}
\gamma^{0}\left[(p_0-\e{p}{-})\Lambda^{+}_{p}
+(p_0+\e{p}{+})\Lambda^{-}_{p}\right] & \Delta \\
\tilde{\Delta} &\gamma^{0}\left[(p_0-\e{p}{+})\Lambda^{+}_{p}
+(p_0+\e{p}{-})\Lambda^{-}_{p}\right] \end{array}\right),
\label{G-inv}
\end{eqnarray}
where $\tilde{\Delta}=\gamma^0\Delta^{\dagger}\gamma^0$,
$\Delta_{ab}^{ij} \equiv \gamma^{5} \varepsilon^{ij} \varepsilon_{ab3}
\left[\Delta^{-} \Lambda^{-}_{p} + \Delta^{+} \Lambda^{+}_{p}\right]$
and the ``on-shell" projectors $\Lambda^{(\pm )}_{p}$  are
\be
\Lambda^{(\pm )}_{p}=\frac{1}{2}
\left( 1 \pm \frac{\vec{\alpha}\cdot \vec{p}}{|\vec{p}|} \right),
\quad \mbox{where} \quad 
\vec{\alpha} = \gamma^{0} \vec{\gamma},
\label{Lambda}
\ee
(note that $\Delta^{\pm}$ are complex valued gap functions). Color and
flavor indices are denoted by small latin letters from the beginning and
the middle of the alphabet, respectively. By definition, $\e{p}{\pm} =
|\vec{p}|\pm \mu$ and $\vpsl = -\vec{p} \cdot\vec{\gamma}$. The effects of
the quark wave function renormalization are neglected here.

Now, after inverting the expression in Eq.~(\ref{G-inv}),
we arrive at the following propagator:
\begin{eqnarray}
G(p)&=&i \left(\begin{array}{cc}
R_{11} & R_{12} \\
R_{21} & R_{22} \end{array}\right),
\label{G}
\end{eqnarray}
where
\begin{mathletters}
\begin{eqnarray}
R_{11}(p)&=& \gamma^{0} {\cal I}_{1} 
\left[ \frac{p_{0}+\e{p}{-}}{E^{-}_{\Delta}} \Lambda^{-}_{p} 
+\frac{p_{0}-\e{p}{+}}{E^{+}_{\Delta}} \Lambda^{+}_{p} \right]
+\gamma^{0} {\cal I}_{2}
\left[ \frac{1}{p_{0}-\e{p}{-}} \Lambda^{-}_{p} 
+\frac{1}{p_{0}+\e{p}{+}} \Lambda^{+}_{p} \right],\\
R_{22}(p)&=& \gamma^{0} {\cal I}_{1}
\left[ \frac{p_{0}+\e{p}{+}}{E^{+}_{\Delta}} \Lambda^{-}_{p}
+\frac{p_{0}-\e{p}{-}}{E^{-}_{\Delta}} \Lambda^{+}_{p} \right]
+\gamma^{0} {\cal I}_{2}
\left[ \frac{1}{p_{0}-\e{p}{+}}\Lambda^{-}_{p}
+\frac{1}{p_{0}+\e{p}{-}}\Lambda^{+}_{p} \right],\\
R_{12}(p)&=& \gamma^{5} \hat{\varepsilon} \left[ 
\frac{\Delta^{+}\Lambda^{-}_{p}}{E^{+}_{\Delta}} 
+\frac{\Delta^{-}\Lambda^{+}_{p}}{E^{-}_{\Delta}} 
\right], \\
R_{21}(p)&=& - \gamma^{5} \hat{\varepsilon} \left[ 
\frac{(\Delta^{-})^{*}\Lambda^{-}_{p}}{E^{-}_{\Delta}} 
+\frac{(\Delta^{+})^{*}\Lambda^{+}_{p}}{E^{+}_{\Delta}} \right],
\end{eqnarray}
\end{mathletters}
with $E^{\pm}_{\Delta}=p_{0}^{2}-(\e{p}{\pm})^{2}-|\Delta^{\pm}|^{2}$,
and the three color-flavor matrices are defined as follows:
\ba
\left({\cal I}_{1}\right)_{ab}^{ij} &=&
\left(\delta_{ab} - \delta_{a3}\delta_{b3}\right) \delta^{ij},\\ 
\left({\cal I}_{2}\right)_{ab}^{ij} &=&
\delta_{a3}\delta_{b3}\delta^{ij}, \\
\hat{\varepsilon}_{ab}^{ij} &=& 2i T^{2}_{ab} \varepsilon^{ij}.
\ea
Notice that, when using this quark propagator in loop calculations, one
should make the substitutions
\ba
E^{\pm}_{\Delta} &\to& E^{\pm}_{\Delta} +i \epsilon, \\
p_0 \pm \e{p}{-} &\to& p_0 \pm \e{p}{-} 
\mp i \epsilon \mbox{~sign}(\e{p}{-}), \\
p_0 \pm \e{p}{+} &\to& p_0 \pm \e{p}{+} 
\mp i \epsilon,
\ea
in the denominators, and take the limit of vanishing $\epsilon$ at the
end. This is important for preserving the causality of the theory.

\section{Description of the formalism}
\label{formalism}

Let us start from a simple observation. If the model under consideration
had real NG bosons in the spectrum, its effective potential as a function
of the order parameter $\Delta^{ij}_{ab}$ would have a degenerate manifold
of minima. The dimension of such a manifold would be equal to the number
of the NG bosons. We know, however, that no global symmetries are broken
in the $S2C$ phase and, therefore, the potential should have a single
non-degenerate global minimum. The existence of the pseudo-NG bosons
means, however, that the potential is nearly degenerate along selected
directions. The curvature along these directions defines the masses of the
pseudo-NG bosons. In the limit of the zero curvature, masses go to zero,
as it should be for the NG bosons.

In order to select the directions in the color-flavor space that
correspond to the five pseudo-NG bosons introduced in
Ref.~\cite{Miransky:2000gg,Miransky:2000jt}, we should recall their
definition. These pseudo-NG bosons correspond to ``breaking" of the
approximate axial color symmetry, given by the following transformations
of the quark fields:
\begin{mathletters}
\ba
\psi_{D} &\to& U {\cal P}_{+} \psi_{D} + U^{\dagger} {\cal P}_{-} \psi_{D}, \\
\bar{\psi}_{D} &\to& \bar{\psi}_{D} {\cal P}_{-} U^{\dagger} 
              +\bar{\psi}_{D} {\cal P}_{+} U, \\
\psi^{C}_{D} &\to& U^{*} {\cal P}_{-} \psi^{C}_{D} 
             +U^{T} {\cal P}_{+} \psi^{C}_{D}, \\
\bar{\psi}^{C}_{D} &\to& \bar{\psi}^{C}_{D} {\cal P}_{+} U^{T}
                 + \bar{\psi}^{C}_{D} {\cal P}_{-} U^{*}.
\ea
\end{mathletters}
Of course, this is {\em not} an exact symmetry of the model.  For example,
the kinetic term of the Lagrangian density transforms as follows:
\be
\bar{\psi}_{D} \left(i\dinot +\mu\gamma^{0} + \Ahatnot \right) \psi_{D} \to 
\bar{\psi}_{D} \left(i\dinot +\mu\gamma^{0} 
+{\cal P}_{+} U \Ahatnot U^{\dagger}
+{\cal P}_{-}U^{\dagger} \Ahatnot U \right)\psi_{D}, 
\ee
and no transformation of the vector field could promote this
transformation to a symmetry.

The axial color transformation, as defined above, allows us to 
explicitly extract the phase factors of the gap that correspond
to the nearly degenerate directions of interest,
\be
\Delta \to {\cal P}_{+} U^{\dagger} \Delta U^{*}
         + {\cal P}_{-} U \Delta U^{T}, \quad 
U=\exp\left(i\omega^{A}T^{A}\right). 
\ee
One could consider $\omega^{A}$ as the dynamical fields of the pseudo-NG
bosons (which, up to a factor of the decay constant, are related to the
canonical fields). Such a substitution leads to the following changes of
the components of the quark propagator:
\begin{mathletters}
\ba 
R_{11} &\to& 
R_{11}(\omega) = {\cal P}_{+} U^{\dagger} R_{11} U 
         + {\cal P}_{-} U R_{11} U^{\dagger} ,
\label{11U}\\ 
R_{22} &\to& 
R_{22}(\omega) = {\cal P}_{-} U^{T} R_{22} U^{*} 
         + {\cal P}_{+} U^{*} R_{22} U^{T} ,
\label{22U}\\
R_{12} &\to& 
R_{12}(\omega) = {\cal P}_{+} U^{\dagger} R_{12} U^{*} 
         + {\cal P}_{-} U R_{12} U^{T} ,
\label{12U}\\
R_{21} &\to& 
R_{21}(\omega) = {\cal P}_{-} U^{T} R_{21} U 
         + {\cal P}_{+} U^{*} R_{21} U^{\dagger} ,
\label{21U}
\ea
\label{transform}
\end{mathletters}
assuming that the fields $\omega^{A}$ are constant in space-time.

\begin{figure}
\epsfbox{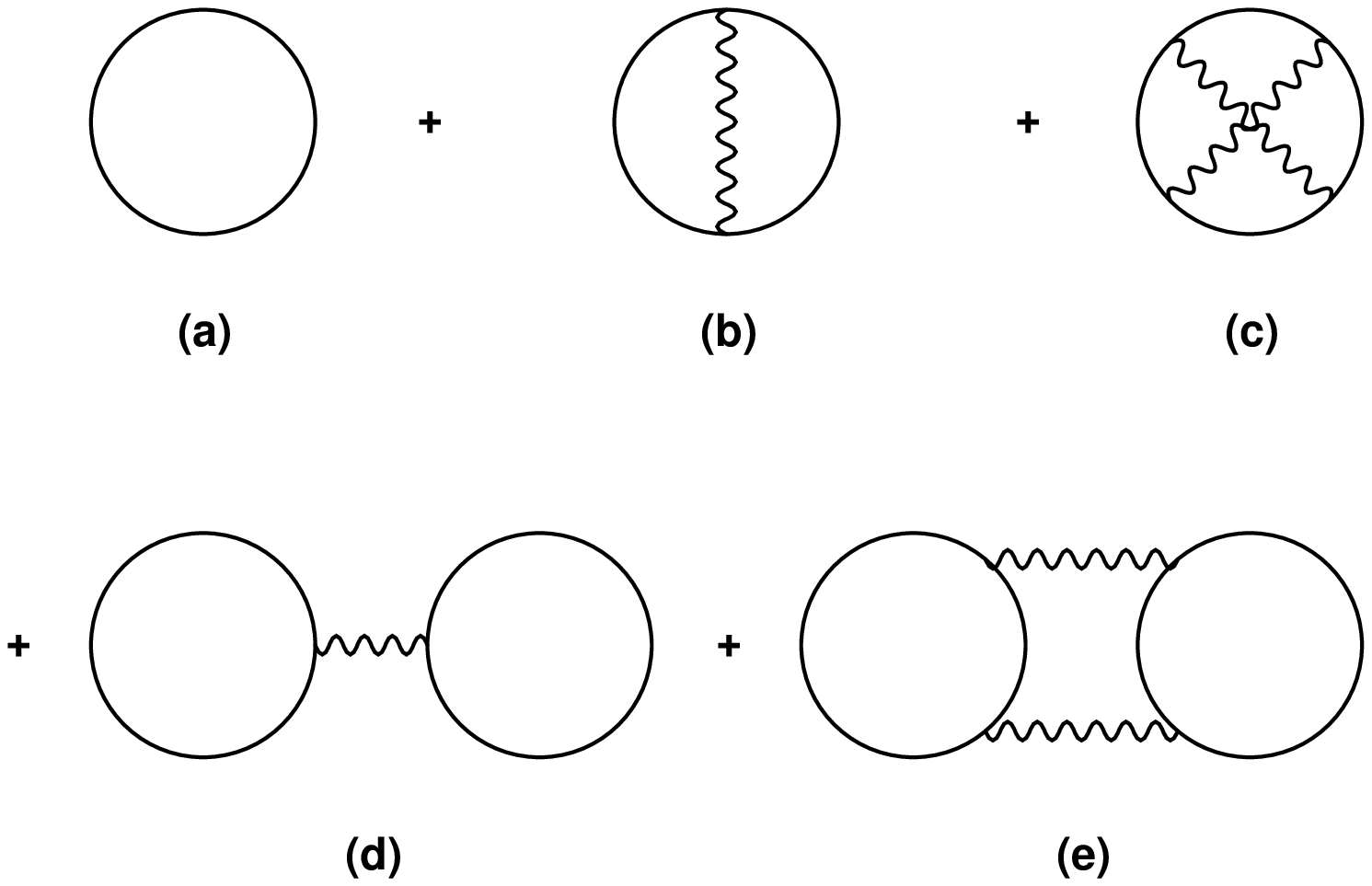}
\caption{Vacuum energy diagrams.}
\label{fig-1}
\end{figure}

In order to construct the effective potential, we use the CJT formalism
\cite{Cornwall:1974vz}. The corresponding general expression reads
\be
V = i \int \frac{d^{4} p}{(2\pi)^{4}} \mbox{Tr} \left[
\ln G(p) S^{-1}(p) - S^{-1}(p) G(p) +1 \right]
+V_{2}[G],
\ee
where $V_{2}[G]$ represents the two-particle irreducible (with respect to
quark lines) contributions (we will discuss this point below).
There are, in general, an infinite number of
diagrams in $V_{2}[G]$. In our analysis, we leave only a few leading order
diagrams, graphically shown in Fig.~\ref{fig-1}.

Before proceeding to the actual calculation, let us try to understand
which type of diagrams could produce a non-trivial dependence of the
potential on the pseudo-NG fields $\omega^{A}$. To this end, we have to
recall the origin of the pseudo-NG bosons under consideration. In
particular, it is crucial that their appearance is related to the breaking
of the approximate axial color symmetry. It is clear, then, that the
dependence of the effective potential on the pseudo-NG boson fields
results from some mixing between the left-handed and the right-handed
sectors of the theory.  Since there is no any left-right mixing in the
diagrams containing a single quark loop [diagrams $(a)$, $(b)$ and $(c)$
in Fig.~\ref{fig-1}], the first three diagrams in Fig.~\ref{fig-1} turn
out to be irrelevant for our calculation. The corresponding contributions
to the effective potential are free of any dependence on the constant
$\omega^{A}$ fields.

Now, we move over to more complicated diagrams $(d)$ and $(e)$ in
Fig.~\ref{fig-1}. Notice, that these last two diagrams are not
two-particle irreducible with respect to gluon lines. This is consistent
with the fact that we consider the CJT action as a functional of only the
quark propagator [for a discussion of this point see Ref. \cite{book}].
Both diagrams could potentially produce non-trivial mass
corrections for pseudo-NG bosons. In diagrams $(d)$ and $(e)$, a
non-trivial mixing of the left- and right-handed quark sectors is possible
because there are two separate quark loops.
 
A simple calculation shows that the diagram in Fig.~\ref{fig-1}d gives no
correction to the effective potential. Thus, the leading order corrections
come from the diagram in Fig.~\ref{fig-1}e. The details of our calculation
are presented in the next section.

\section{Leading order calculations}
\label{vac-diagram}

As we argued in the preceding section, the leading order corrections to
the masses of the pseudo-NG bosons come from the diagram in
Fig.~\ref{fig-1}e. In this section, we give the details of the calculation
and derive an approximate analytical result for the masses.

The analytical expression for the vacuum diagram in Fig.~\ref{fig-1}e
reads
\ba
V(\mbox{Fig.}~\ref{fig-1}\mbox{e}) &=& - 2i \pi^{2} \alpha_{s}^{2} \int 
\frac{ d^{4} p d^{4} k d^{4} q}{(2\pi)^{12}} 
\mbox{Tr}\left[ \Gamma^{A\mu} G(p) \Gamma^{B\kappa} G(p-q) \right]
\mbox{Tr}\left[ \Gamma^{A\nu} G(k-q) \Gamma^{B\lambda} G(k) \right]
D_{\mu\nu} (q) D_{\kappa\lambda} (q),
\label{V}
\ea
where the vertex is
\be
\Gamma^{A\mu} = \gamma^{\mu} \left( \begin{array}{ll}
T^{A} & 0 \\
0 & -\left( T^{A} \right)^{T}
\end{array} \right).
\ee
The expression in Eq.~(\ref{V}) contains two factors of the following
type:
\be
2i \pi \alpha_{s}\int\frac{d^{4}p}{(2\pi)^{4}}
\mbox{Tr}\left[ \Gamma^{A\mu} G(p)
\Gamma^{B\kappa} G(p-q) \right].
\label{block}
\ee
In order to extract the dependence of this quantity on the psedo-NG boson
fields $\omega^{A}$, we perform the substitutions of all component
functions, given in Eq.~(\ref{transform}). At the end, we expand
the result in powers of $\omega^{A}$, keeping the terms up to the second
order. Thus, we arrive at the following result:
\ba
&& 2i \pi \alpha_{s}\int\frac{d^{4}p}{(2\pi)^{4}}
\mbox{tr}\Bigg[
{\cal P}_{+} \gamma^{\mu} \tilde{T}^{A}_{\omega} R_{12}(p)
\gamma^{\kappa} \left(\tilde{T}^{B}_{\omega}\right)^{T} R_{21}(p-q)
+{\cal P}_{-} \gamma^{\mu} T^{A}_{\omega} R_{12}(p)
\gamma^{\kappa} \left(T^{B}_{\omega}\right)^{T} R_{21}(p-q)
\nonumber \\
&+&{\cal P}_{+} \gamma^{\mu} \left(T^{A}_{\omega}\right)^{T}
R_{21}(p) \gamma^{\kappa} T^{B}_{\omega} R_{12}(p-q)
+{\cal P}_{-} \gamma^{\mu} \left(\tilde{T}^{A}_{\omega}\right)^{T}
R_{21}(p) \gamma^{\kappa} \tilde{T}^{B}_{\omega} R_{12}(p-q)
\nonumber \\  
&-&{\cal P}_{+} \gamma^{\mu} \tilde{T}^{A}_{\omega} R_{11}(p)
\gamma^{\kappa} \tilde{T}^{B}_{\omega} R_{11}(p-q)
-{\cal P}_{-} \gamma^{\mu} T^{A}_{\omega} R_{11}(p)
\gamma^{\kappa} T^{B}_{\omega} R_{11}(p-q)
\nonumber \\
&-&
{\cal P}_{+} \gamma^{\mu} \left(T^{A}_{\omega}\right)^{T}
R_{22}(p) \gamma^{\kappa} \left(T^{B}_{\omega}\right)^{T}
R_{22}(p-q)
-{\cal P}_{-} \gamma^{\mu} \left(\tilde{T}^{A}_{\omega}\right)^{T}
R_{22}(p) \gamma^{\kappa} \left(\tilde{T}^{B}_{\omega}\right)^{T}
R_{22}(p-q)
\Bigg] \nonumber \\
&=& \Pi^{AB,\mu\kappa}(q)
+ \omega^{X} \omega^{Y} f^{XAC} f^{YBD} \Pi^{CD,\mu\kappa}(q)
\nonumber \\
&+& \frac{1}{2} \omega^{X} \omega^{Y} f^{XAD} f^{YDC}
\Pi^{CB,\mu\kappa}(q)
+ \frac{1}{2} \omega^{X} \omega^{Y} f^{XBD} f^{YDC}
\Pi^{AC,\mu\kappa}(q).
\label{expansion}
\ea
Here we used the shorthand notation,
\begin{mathletters}
\ba
T^{A}_{\omega} \equiv U T^{A} U^{\dagger}
&\simeq & T^{A} - \omega^{B} f^{BAC} T^{C}
+\frac{1}{2} \omega^{B} \omega^{C} f^{BAD} f^{CDE} T^{E}
+ \ldots, \\
\tilde{T}^{A}_{\omega} \equiv U^{\dagger} T^{A} U
&\simeq& T^{A} + \omega^{B} f^{BAC} T^{C}
+\frac{1}{2} \omega^{B} \omega^{C} f^{BAD} f^{CDE} T^{E}
+ \ldots,
\ea
\end{mathletters}
where $f^{BAD}$ are the structure constants of $SU(3)$.
Also, we introduced the one loop polarization tensor in the color
superconducting phase \cite{Rischke:2000qz},
\ba
\Pi^{AB,\mu\kappa}(q)&=& 2i \pi \alpha_{s}\int\frac{d^{4}p}{(2\pi)^{4}}
\mbox{tr} \Bigg[ 
\gamma^{\mu} T^{A} R_{12}(p) \gamma^{\kappa} \left(T^{B}\right)^{T} 
R_{21}(p-q)
+\gamma^{\mu} \left(T^{A}\right)^{T} R_{21}(p) \gamma^{\kappa} T^{B} 
R_{12}(p-q) \nonumber\\
&-&\gamma^{\mu} T^{A} R_{11}(p) \gamma^{\kappa} T^{B} R_{11}(p-q)
-\gamma^{\mu} \left(T^{A}\right)^{T} R_{22}(p) \gamma^{\kappa}
\left(T^{B}\right)^{T} R_{22}(p-q) \Bigg].
\ea
By performing the traces over the color and flavor indices, we 
arrive at the following expression for the polarization tensor
(see Appendices~\ref{pol-tensor} and \ref{integrals} for details):
\begin{mathletters}
\ba
&&\left. \Pi^{AB,\mu\nu}(q)\right|_{A,B=1,2,3}
= \delta^{AB} \Pi^{\mu\nu}_{1}(q),\\
&&\left. \Pi^{AB,\mu\nu}(q)\right|_{A,B=4,5,6,7}
= \delta^{AB} \Pi^{\mu\nu}_{4}(q)
+i\left( \delta^{A4}\delta^{B5} -\delta^{A5}\delta^{B4}
+\delta^{A6}\delta^{B7}-\delta^{A7}\delta^{B6}
\right) \tilde{\Pi}_{4}^{\mu\kappa}(q),\\
&&\Pi^{88,\mu\nu}(q)= \Pi^{\mu\nu}_{8}(q).
\ea
\end{mathletters}
After substituting the expansion (\ref{expansion}) in Eq.~(\ref{V}), 
we arrive at 
\ba
V(\mbox{Fig.}~\ref{fig-1}\mbox{e}) &=& \frac{i}{2} \int
\frac{d^{4} q}{(2\pi)^{4}}
\Pi^{AB,\mu\kappa}(q) \Pi^{AB,\nu\lambda}(q)
D_{\mu\nu} (q) D_{\kappa\lambda} (q) \nonumber \\
&-& \frac{3i}{4} \sum_{A=4}^{7} (\omega^{A})^{2}
\int \frac{d^{4} q}{(2\pi)^{4}}
D_{\mu\nu} (q) D_{\kappa\lambda} (q) \nonumber \\
&\times& \Bigg\{ 
\left[ \Pi_{4}^{\mu\kappa}(q)-\Pi_{1}^{\mu\kappa}(q)\right]
\left[ \Pi_{4}^{\nu\lambda}(q)-\Pi_{1}^{\nu\lambda}(q)\right]
+2 \tilde{\Pi}_{4}^{\mu\kappa}(q) \tilde{\Pi}_{4}^{\nu\lambda}(q)
\nonumber \\
&+&\left[ \Pi_{4}^{\mu\kappa}(q)-\Pi_{8}^{\mu\kappa}(q)\right]
\left[ \Pi_{4}^{\nu\lambda}(q)-\Pi_{8}^{\nu\lambda}(q)\right]
\Bigg\}+ O\left[(\omega^{A})^{4}\right] .
\label{V-4}
\ea
It is noticeable that the right hand side of the last expression is
independent of the $\omega^{8}$ field, related to the singlet pseudo-NG
boson. This means that the diagram in Fig.~\ref{fig-1}e does not give any
non-trivial contribution to the value of the corresponding mass.  Only the
(anti-)doublet pseudo-NG bosons get a non-zero mass.

In order to understand this point, it is instructive to consider the
``ideal" case, assuming that the axial color symmetry generated by
$\gamma^{5}T^{8}$ is a true (rather than approximate) symmetry of dense
QCD. Then, the singlet pseudo-NG boson would be related to the breaking of
the restricted axial symmetry, defined by the following transformation:
\begin{mathletters}
\ba
\psi &\to& \exp\left(i\omega\gamma^{5}\right) {\cal I}_{1} \psi
             + {\cal I}_{2} \psi, \\
\bar{\psi} &\to& \bar{\psi} {\cal I}_{1}
\exp\left(i\omega\gamma^{5}\right)+\bar{\psi} {\cal I}_{2}, \\
\psi^{C} &\to& \exp\left(i\omega\gamma^{5}\right) {\cal I}_{1} \psi^{C}
             +{\cal I}_{2} \psi^{C}, \\
\bar{\psi^{C}} &\to& \bar{\psi^{C}} {\cal I}_{1}
\exp\left(i\omega\gamma^{5}\right)
                 + \bar{\psi^{C}} {\cal I}_{2}.
\ea
\end{mathletters}
acting on the first two color quarks. The other symmetry
[$\tilde{U}_{5}(1)$], acting on the third color quarks would remain
unbroken. This is a simple consequence of the fact that the third color
quarks do not participate in the color condensation. Now, the axial color
transformation generated by $\gamma^{5}T^{8}$ could be thought of as the
ordinary axial transformation $U_{5}(1)$ accompanied by the unbroken
$\tilde{U}_{5}(1)$. Therefore, it appears to be equivalent to say that
either the restricted or the ordinary axial symmetry is spontaneously
broken in this ideal limit. In reality, both the restricted and the
ordinary axial symmetries are explicitly broken. However, while the former
is broken perturbatively, the latter is broken by much smaller
non-perturbative (instanton-like) effects. Because of that, the singlet
pseudo-NG boson is connected with the ordinary axial transformation
$U_{5}(1)$ and its mass is zero in any order of the expansion in the
coupling constant.

Non-perturbative analysis reveals that the singlet pseudo-NG boson
$\eta$ has a non-zero mass. The value of the mass was estimated in
Ref.~\cite{Son:2000fh}. In our notation, it reads
\be
M^{2}_{\eta} \simeq \frac{C_{\eta}}{\alpha_{s}^{7}}
|\Delta^{-}|^{2} \exp\left(-\frac{2\pi}{\alpha_{s}}\right), \qquad
C_{\eta} \simeq 10^{5}.
\ee
The parametric dependence of this mass of the $\eta$ pseudo-NG singlet
clearly indicates the non-perturbative nature of the underlying dynamics.
To derive this estimate, one needs to consider the instanton contribution
(notice the characteristic exponential factor in the expression above)
to the vacuum energy in the color superconducting phase.

Before calculating the masses of the other pseudo-NG bosons, let us turn
to the question of the reliability (and, in particular, gauge invariance)
of the loop expansion for the CJT effective potential in the present
problem. From Eq.~(\ref{V-4}) we see that the vacuum energy is expressed
through the one-loop polarization tensor. As is well known, the Ward
(Slavnov-Taylor) identities (or, equivalently, the BRST transformations)
imply that in both Abelian and non-Abelian gauge field theories, the
polarization tensor is transverse in covariant gauges. We show in
Appendix~\ref{tra-pol-ten}, however, that while the one-loop polarization
tensor connected with the unbroken $SU(2)_{c}$ is indeed transverse [see
Eqs.~(\ref{Pi_1-IR}), (\ref{Pi_1-general}) and (\ref{Pi_1-gen})], the
one-loop polarization tensors connected with the five broken generators
are not [see Eqs.~(\ref{Pi_48-IR}), (\ref{Pi_4-general}) and
(\ref{Pi_8-general}) in Appendix~\ref{tra-pol-ten}].

What is the reason for the violation of the Ward identities in the
one-loop approximation? The answer is that it is connected with the
Meissner-Higgs effect. As was already emphasized in our paper
\cite{Miransky:2000jt} and in the earlier papers
\cite{Jackiw:1973tr,Cornwall:1973ts}, this effect implies the presence of
unphysical composites having quantum numbers of the (would be) NG bosons
in any non-unitary gauge, including of course all covariant ones. Although
in the unitary gauge these composites are ``eaten" by the five massive
gluons, in other gauges they are crucial for preserving both the unitarity
and the Ward identities (i.e. gauge invariance).\footnote{Note that,
because of the composite (diquark) nature of the order parameter in color
superconductivity, it does not seem to be straightforward to define and to
use the unitary gauge in this case.} In particular, they lead to an
important pole correction in the quark-gluon vertex function
\cite{Miransky:2000jt}.

Actually, the traces of this problem are known to appear in other studies
in dense QCD. For example, it shows up even in the analysis of the
Schwinger-Dyson equation in the hard dense loop (HDL) improved ladder
approximation \cite{Hong:2000fh,Rajagopal:2000rs}. In this case, the gauge
dependence of the gap appears only through a dependence in the
preexponential factor of the gap. Formally, it is the
next-to-next-to-leading order correction. Usually, therefore, this gauge
dependence of the superconducting gap is ignored without even studying its
real origin. While such an attitude is quite harmless in the case of the
Schwinger-Dyson equation, it turns out to be crucial in the present
problem of calculating the masses of the pseudo-NG bosons. Indeed, as is
shown in Appendix~\ref{tra-pol-ten}, the gauge dependence of their masses,
caused by the non-transversality of the polarization tensor, is strong.
The reason is that while in the Schwinger-Dyson equation the dominant
region is that with momenta much larger than the fermion gap
$|\Delta^{-}|$, in this problem the infrared region with momenta less than
$|\Delta^{-}|$ dominates.

This implies that a consistent approximation for calculating the CJT
potential has to include those (would-be) NG composites. This in turn
implies that one should modify the ordinary loop expansion of the CJT
potential. Since here we are interested in the infrared dynamics, one can
neglect the composite structure of these bosons.  This leads to a
non-linear realization of color symmetry breaking.  In particular, one has
to calculate the polarization tensor in this framework.

In Appendix~\ref{tra-pol-ten}, we consider this problem. As is shown
there, including the (would be) NG bosons indeed restores the
transversality of the polarization tensor.  The remarkable thing is that
the dominant contribution to the masses of the pseudo-NG bosons comes only
from the part connected with the magnetic gluons [singled out by the
projection operator $O^{(1)}$, see Eq.~(\ref{Pi-i-decomp})], and this
part, unlike the other contributions, remains unchanged by including the
contribution of the would be NG composites. This implies that for the
calculation of the masses of pseudo-NG bosons in the Landau gauge one can
use the initial, unmodified framework for the CJT effective potential (at
least in this approximation). Indeed, the dangerous longitudinal terms in
the polarization operator does not contribute in this gauge and only the
contribution of the magnetic gluons matters. Because of this observation,
henceforth we will use this gauge.

To get a rough estimate of the mass of doublets, one could use the
following order of magnitude asymptotes for the polarization tensor
[see Eqs.~(\ref{B8}), (\ref{B16}), (\ref{B21}) and (\ref{B24}) in
Appendix~\ref{integrals} and Eq. (\ref{Pi-i-decomp}) in Appendix
\ref{tra-pol-ten}]:
\ba
\Pi_{4}^{(1)}(q) - \Pi_{1}^{(1)}(q) &\sim &
\Pi_{4}^{(1)}(q) - \Pi_{8}^{(1)}(q) \sim
\left\{ \begin{array}{lll}
\alpha_{s} \mu^2 ,
& \mbox{for} & |q_{4}|,q \ll 2|\Delta^{-}|; \\
\frac{\alpha_{s} \mu^2|\Delta^{-}|^{2}}{q_{4}^{2}}
\ln\frac{q_{4}^{2}}{|\Delta^{-}|^{2}},
& \mbox{for} & q, |\Delta^{-}| \ll |q_{4}|; \\  
\frac{\alpha_{s} \mu^2|\Delta^{-}|^{2}}{|q_{4}| q},
& \mbox{for} & |\Delta^{-}| \ll |q_{4}| \ll q;\\
\frac{\alpha_{s} \mu^2|\Delta^{-}|}{q},
& \mbox{for} &  |q_{4}| \ll |\Delta^{-}| \ll q.  
\end{array}\right.
\ea
By making use of these asymptotes along with the HDL expression for the
gluon propagator [see, for example, Ref.~\cite{Hong:2000fh}], we check
that the dominant contribution to the quadratic term in Eq.~(\ref{V-4})  
in the Landau gauge comes from the region of momenta where $|\Delta| \alt
q \alt \mu$ and $0 \alt |q_{4}| \alt |\Delta|$. Therefore, by taking the 
explicit expressions for the polarization tensor given in 
Appendix~\ref{tra-pol-ten} into account, we derive
\ba
M^2 &= & \frac{1}{F^{2}} \frac{\partial^{2} V}{\partial \omega^{2}}
\simeq \frac{170\pi}{3\mu^{2}} \int_{|\Delta|}^{\mu} q^{2}
d q \int_{0}^{|\Delta|} d q_{4}
\frac{q^2}{\left(q^3 +\frac{\pi}{2} m_{D}^{2} |q_{4}| \right)^{2}}
\left(\frac{\alpha_{s} \mu^{2}|\Delta|}{q} \right)^{2}\nonumber \\
&\simeq&  \frac{340\pi}{9} \alpha_{s} |\Delta|^{2}
\ln\frac{\mu}{|\Delta|} 
\simeq C_{M} \sqrt{\alpha_{s}} |\Delta|^{2}, 
\quad C_{M} \approx  \frac{85\pi^{2}}{3} \sqrt{2 \pi} 
\approx 7\times 10^{2},
\label{LOresult}
\ea
where the definition of the Debye mass, $m_{D}^{2} = (2/\pi) \alpha_{s}
\mu^2$, the decay constant $F^{2} = \mu^2/8\pi^{2}$, as well as the
relation between the gap and the chemical potential,
$\ln(\mu/|\Delta|) \simeq 3 (\pi/2)^{3/2}\alpha_{s}^{-1/2}$, 
were used.

\section{Pseudo-NG (anti-) doublets versus confinement}
\label{2-confinement}

In the color superconducting phase in the model with two quark flavors,
the $SU(2)_{c}$ subgroup remains unbroken. Since the only massless quarks
of the third color do not interact with gluons of $SU(2)_{c}$, the
corresponding dynamics of gluons decouples in the far infrared region, 
$p \ll |\Delta^{-}|$. The low energy effective action of the $SU(2)_{c}$
gluondynamics was derived in Ref.~\cite{Rischke:2000cn}. Of special
interest for us here is the observation of Ref.~\cite{Rischke:2000cn} that
the unbroken $SU(2)_{c}$ is confined at sufficiently low energies. The
corresponding scale was also estimated. It is given by the following
simple expression:
\be
\Lambda_{QCD}^{\prime}\simeq |\Delta^{-}|\exp
\left[-C_{0}\alpha_{s}^{2}\exp\left(C/\sqrt{\alpha_{s}}\right)\right], 
\qquad
C=3\left(\frac{\pi}{2}\right)^{3/2},
\label{LambdaQCD}
\ee
(there is some uncertainty in determining $C_{0}\simeq 10^{-3}$).

Now, let us discuss how such confinement could affect the properties of
the pseudo-NG bosons. In particular, this concerns the doublet and
antidoublet states which are colored under the leftover $SU(2)_{c}$
subgroup. Because of the confinement, these pseudo-NG states cannot exist
as free particles. Instead, they should form some colorless bound states.

The colorless bound states should be somewhat similar to charmonium in
ordinary QCD. In fact, the situation with the doublet-antidoublet states
in the color superconducting phase is much simpler. This is because the
new confinement scale $\Lambda_{QCD}^{\prime}$ is extremely small as
compared to the masses of the (anti-) doublets themselves. This already
suggests that the binding dynamics is essentially perturbative
[this could also be checked posterior, see Eq.~(\ref{binding-e}) below].
The bound colorless states would be similar to the positronium in 
QED \cite{Gross:1969rv,Barbieri:1978mf,Lepage:1977gd}.
The only difference is due to the color superconducting medium 
effects. The Coulomb potential between two static charges
at a distance $r$ from each other is given by \cite{Rischke:2000cn}
\be
V_{C} \simeq \frac{4 \pi \alpha_{s}}{\epsilon r}, \qquad 
\frac{1}{|\Delta^{-}|} \ll r \ll \frac{1}{\Lambda_{QCD}^{\prime}}, 
\ee 
where 
\be
\epsilon \approx \frac{2 \alpha_{s} \mu^{2}}{9 \pi |\Delta^{-}|^{2}}
\label{diel-constant}
\ee
is the dielectric constant \cite{Rischke:2000cn} [see also our 
derivation of Eq.~(\ref{Pi_1-IR}) in Appendix~\ref{tra-pol-ten}]. 
By analogy with the positronium (notice that the constituent 
doublets are spinless in the problem at hand, but such a difference 
affects only the fine structure of the spectrum),
we get the following estimate for the binding energies of the 
colorless states built of the doublet-antidoublet pairs:
\be
E_{n} \simeq -\frac{M \alpha_{s}^{2}}{4 \epsilon^{2} n^{2}} 
\simeq C_{E} \frac{\alpha_{s}^{-39/4} |\Delta^{-}|}{n^{2}} 
\exp\left(-\frac{3 \sqrt{2} \pi^{3/2} }{\sqrt{\alpha_{s}}}
\right), \qquad n=1,2,\ldots,
\label{binding-e}
\ee
with $C_{E} \approx 1.8 \times 10^{11}$. Notice that this binding energy
is parametrically much larger than the confinement scale
$\Lambda_{QCD}^{\prime}$ in Eq.~(\ref{LambdaQCD}) when $\alpha_{s} \to 0$.
This means that it is the (perturbative) Coulomb contribution to the
potential that is mostly responsible for the paring dynamics of the
(anti-) doublets into colorless hadrons.  The linear confining
(non-perturbative) part of the potential is of minor importance at least
for the low lying states.

We recall that in the $S2C$ phase, there exists a conserved (modified)
baryon charge. While the value of this charge equals zero for the quarks
of the first two colors, it is $+1$ for the third color (massless) quarks.
Since the pseudo-NG (anti-) doublets are composed of one massive (anti-)
quark and one massless (anti-) quark, they are scalar baryons, carrying
the baryon charge $\pm{1}$. The colorless hadrons composed of them
(discussed above) can be called mesons.

\section{Conclusion}
\label{conclusion}

In this paper, we derived analytical estimates for the masses of the
pseudo-Nambu-Goldstone bosons in the color superconducting phase of dense
QCD with two light flavors. In agreement with our previous hypothesis, the
mass values are small compared to the value of the superconducting gap at
sufficiently large quark densities. Analytically, the expression for the
masses of doublet and antidoublet pseudo-NG bosons are suppressed by a
power of the coupling constant. The mass of the singlet is exponentially
suppressed. The mechanism for mass generation of the singlet is purely
non-perturbative \cite{Son:2000fh}.

As was shown in \cite{Rischke:2000cn}, the unbroken $SU(2)_{c}$ subgroup
is subject to confinement. This fact has an important effect on the
properties of the doublet and antidoublet pseudo-NG bosons which are
colored with respect to $SU(2)_{c}$. In particular, we argue that the
doublets and antidoublets should form charmonium-like colorless bound
states. Moreover, since the confinement scale of $SU(2)_{c}$ subgroup is
extremely small compared to the value of the masses of doublet and
antidoublet, the pairing dynamics of the colorless mesons is dominated by
perturbative Coulomb-like forces. This allows us to estimate
the corresponding binding energy.

\begin{acknowledgments}

V.A.M. and I.A.S. would like to thank V.P. Gusynin for many useful
discussions. I.A.S. also thanks J.A.~Bowers, D.~Rishcke, D.T.~Son,
M.~Stephanov and I.~Zahed for comments and discussions at QM'2001.
The work of V.A.M. was partially supported by Grant-in-Aid of Japan
Society for the Promotion of Science (JSPS) No.~11695030.
The work of I.A.S. was supported by the U.S. Department of Energy Grants
No.~DE-FG02-87ER40328. The work of L.C.R.W. was supported by the U.S.
Department of Energy Grant No.~DE-FG02-84ER40153.

\end{acknowledgments}

\appendix

\section{General form of the polarization tensor}
\label{pol-tensor}

The general form of the polarization tensor in the color superconducting
phase was given in \cite{Rischke:2000qz}. For completeness of the
presentation, we also rederive it here
\ba
&& \left. \Pi^{AB,\mu\nu}(q) \right|_{A,B=1,2,3}
\equiv \delta^{AB} \Pi_{1}^{\mu\nu}(q)
= 2i \delta^{AB} \pi \alpha_{s}\int\frac{d^{4}p}{(2\pi)^{4}}
\nonumber \\
&\times& \mbox{tr}_{D}\Bigg[ \gamma^{\mu} \left(
\frac{\Delta^{+}\Lambda^{-}_{p}}{E^{+}_{\Delta}(p)}
+\frac{\Delta^{-}\Lambda^{+}_{p}}{E^{-}_{\Delta}(p)}
\right) \gamma^{\nu} \left(
\frac{(\Delta^{-})^{*}\Lambda^{-}_{p-q}}{E^{-}_{\Delta}(p-q)}
+\frac{(\Delta^{+})^{*}\Lambda^{+}_{p-q}}{E^{+}_{\Delta}(p-q)}
\right) \nonumber \\
&+& \gamma^{\mu} \left(
\frac{(\Delta^{-})^{*}\Lambda^{-}_{p}}{E^{-}_{\Delta}(p)}
+\frac{(\Delta^{+})^{*}\Lambda^{+}_{p}}{E^{+}_{\Delta}(p)}
\right) \gamma^{\nu} \left(
\frac{\Delta^{+}\Lambda^{-}_{p-q}}{E^{+}_{\Delta}(p-q)}
+\frac{\Delta^{-}\Lambda^{+}_{p-q}}{E^{-}_{\Delta}(p-q)}
\right) \nonumber \\
&+& \gamma^{\mu} \gamma^{0} \left(
\frac{p_{0}+\e{p}{-}}{E^{-}_{\Delta}(p)} \Lambda^{-}_{p}
+\frac{p_{0}-\e{p}{+}}{E^{+}_{\Delta}(p)} \Lambda^{+}_{p}
\right) \gamma^{\nu }\gamma^{0} \left(
\frac{p_{0}-q_{0}+\e{p-q}{-}}{E^{-}_{\Delta}(p-q)}
\Lambda^{-}_{p-q}
+\frac{p_{0}-q_{0}-\e{p-q}{+}}{E^{+}_{\Delta}(p-q)}
\Lambda^{+}_{p-q}
\right)\nonumber \\
&+& \gamma^{\mu} \gamma^{0} \left(
\frac{p_{0}+\e{p}{+}}{E^{+}_{\Delta}(p)} \Lambda^{-}_{p}
+\frac{p_{0}-\e{p}{-}}{E^{-}_{\Delta}(p)} \Lambda^{+}_{p}
\right) \gamma^{\nu} \gamma^{0} \left(
\frac{p_{0}-q_{0}+\e{p-q}{+}}{E^{+}_{\Delta}(p-q)}
\Lambda^{-}_{p-q}
+\frac{p_{0}-q_{0}-\e{p-q}{-}}{E^{-}_{\Delta}(p-q)}
\Lambda^{+}_{p-q}
\right)
\Bigg],
\label{Pi-123}
\ea
\ba
&& \left. \Pi^{AB,\mu\nu}(q) \right|_{A,B=4,5,6,7}
\equiv \delta^{AB} \Pi_{4}^{\mu\nu}(q)
+i\left( \delta^{A4}\delta^{B5} -\delta^{A5}\delta^{B4}
+\delta^{A6}\delta^{B7}-\delta^{A7}\delta^{B6}
\right) \tilde{\Pi}_{4}^{\mu\nu}(q)
= i \pi \alpha_{s} \delta^{AB}
\nonumber \\
&\times& \int\frac{d^{4}p}{(2\pi)^{4}} \mbox{tr}_{D}\Bigg[
\gamma^{\mu} \gamma^{0} \left(
\frac{p_{0}+\e{p}{-}}{E^{-}_{\Delta}(p)} \Lambda^{-}_{p}
+\frac{p_{0}-\e{p}{+}}{E^{+}_{\Delta}(p)} \Lambda^{+}_{p}
\right) \gamma^{\nu }\gamma^{0}  \left(
\frac{1}{p_{0}-q_{0}-\e{p-q}{-}} \Lambda^{-}_{p-q}
+\frac{1}{p_{0}-q_{0}+\e{p-q}{+}} \Lambda^{+}_{p-q}
\right) \nonumber \\
&+& \gamma^{\mu} \gamma^{0} \left(
\frac{1}{p_{0}-\e{p}{-}} \Lambda^{-}_{p}
+\frac{1}{p_{0}+\e{p}{+}} \Lambda^{+}_{p} \right)
\gamma^{\nu} \gamma^{0}  \left(
\frac{p_{0}-q_{0}+\e{p-q}{-}}{E^{-}_{\Delta}(p-q)}
\Lambda^{-}_{p-q}
+\frac{p_{0}-q_{0}-\e{p-q}{+}}{E^{+}_{\Delta}(p-q)}
\Lambda^{+}_{p-q}
\right) \nonumber \\
&+& \gamma^{\mu} \gamma^{0} \left(
\frac{p_{0}+\e{p}{+}}{E^{+}_{\Delta}(p)} \Lambda^{-}_{p}
+\frac{p_{0}-\e{p}{-}}{E^{-}_{\Delta}(p)} \Lambda^{+}_{p}
\right) \gamma^{\nu }\gamma^{0}  \left(
\frac{1}{p_{0}-q_{0}-\e{p-q}{+}} \Lambda^{-}_{p-q}
+\frac{1}{p_{0}-q_{0}+\e{p-q}{-}} \Lambda^{+}_{p-q}
\right) \nonumber \\
&+& \gamma^{\mu} \gamma^{0} \left(
\frac{1}{p_{0}-\e{p}{+}} \Lambda^{-}_{p}
+\frac{1}{p_{0}+\e{p}{-}} \Lambda^{+}_{p} \right)
\gamma^{\nu} \gamma^{0}  \left(
\frac{p_{0}-q_{0}+\e{p-q}{+}}{E^{+}_{\Delta}(p-q)}
\Lambda^{-}_{p-q}
+\frac{p_{0}-q_{0}-\e{p-q}{-}}{E^{-}_{\Delta}(p-q)}
\Lambda^{+}_{p-q}
\right)
\Bigg] \nonumber\\
&-& \frac{\pi}{2} \alpha_{s} \left(
\delta^{A4}\delta^{B5} -\delta^{A5}\delta^{B4}
+\delta^{A6}\delta^{B7}-\delta^{A7}\delta^{B6}
\right)
\nonumber \\
&\times&  \int \frac{d^{4} p}{(2\pi)^{4}}\mbox{tr}_{D}\Bigg[
\gamma^{\mu} \gamma^{0} \left(
\frac{p_{0}+\e{p}{-}}{E^{-}_{\Delta}(p)} \Lambda^{-}_{p}
+\frac{p_{0}-\e{p}{+}}{E^{+}_{\Delta}(p)} \Lambda^{+}_{p}
\right) \gamma^{\nu }\gamma^{0}  \left(
\frac{1}{p_{0}-q_{0}-\e{p-q}{-}} \Lambda^{-}_{p-q}
+\frac{1}{p_{0}-q_{0}+\e{p-q}{+}} \Lambda^{+}_{p-q}
\right) \nonumber \\
&-& \gamma^{\mu} \gamma^{0} \left(
\frac{1}{p_{0}-\e{p}{-}} \Lambda^{-}_{p}
+\frac{1}{p_{0}+\e{p}{+}} \Lambda^{+}_{p} \right)
\gamma^{\nu} \gamma^{0}  \left(
\frac{p_{0}-q_{0}+\e{p-q}{-}}{E^{-}_{\Delta}(p-q)}
\Lambda^{-}_{p-q}
+\frac{p_{0}-q_{0}-\e{p-q}{+}}{E^{+}_{\Delta}(p-q)}
\Lambda^{+}_{p-q}
\right) \nonumber \\
&+& \gamma^{\mu} \gamma^{0} \left(
\frac{p_{0}+\e{p}{+}}{E^{+}_{\Delta}(p)} \Lambda^{-}_{p}
+\frac{p_{0}-\e{p}{-}}{E^{-}_{\Delta}(p)} \Lambda^{+}_{p}
\right) \gamma^{\nu }\gamma^{0}  \left(
\frac{1}{p_{0}-q_{0}-\e{p-q}{+}} \Lambda^{-}_{p-q}
+\frac{1}{p_{0}-q_{0}+\e{p-q}{-}} \Lambda^{+}_{p-q}
\right) \nonumber \\
&-& \gamma^{\mu} \gamma^{0} \left(
\frac{1}{p_{0}-\e{p}{+}} \Lambda^{-}_{p}
+\frac{1}{p_{0}+\e{p}{-}} \Lambda^{+}_{p} \right)
\gamma^{\nu} \gamma^{0}  \left(
\frac{p_{0}-q_{0}+\e{p-q}{+}}{E^{+}_{\Delta}(p-q)}
\Lambda^{-}_{p-q}
+\frac{p_{0}-q_{0}-\e{p-q}{-}}{E^{-}_{\Delta}(p-q)}
\Lambda^{+}_{p-q}
\right) \Bigg],
\label{Pi-4567}
\ea
\ba
&& \Pi^{88,\mu\nu}(q) \equiv \Pi_{8}^{\mu\nu} (q)
=- \frac{2}{3} i \pi \alpha_{s}\int\frac{d^{4}p}{(2\pi)^{4}}
\nonumber \\
&\times& \mbox{tr}_{D}\Bigg[ \gamma^{\mu} \left(
\frac{\Delta^{+}\Lambda^{-}_{p}}{E^{+}_{\Delta}(p)}
+\frac{\Delta^{-}\Lambda^{+}_{p}}{E^{-}_{\Delta}(p)}
\right) \gamma^{\nu} \left(
\frac{(\Delta^{-})^{*}\Lambda^{-}_{p-q}}{E^{-}_{\Delta}(p-q)}
+\frac{(\Delta^{+})^{*}\Lambda^{+}_{p-q}}{E^{+}_{\Delta}(p-q)}
\right) \nonumber \\
&+& \gamma^{\mu} \left(
\frac{(\Delta^{-})^{*}\Lambda^{-}_{p}}{E^{-}_{\Delta}(p)}
+\frac{(\Delta^{+})^{*}\Lambda^{+}_{p}}{E^{+}_{\Delta}(p)}
\right) \gamma^{\nu} \left(
\frac{\Delta^{+}\Lambda^{-}_{p-q}}{E^{+}_{\Delta}(p-q)}
+\frac{\Delta^{-}\Lambda^{+}_{p-q}}{E^{-}_{\Delta}(p-q)}
\right) \nonumber \\
&-& \gamma^{\mu} \gamma^{0} \left(
\frac{p_{0}+\e{p}{-}}{E^{-}_{\Delta}(p)} \Lambda^{-}_{p}
+\frac{p_{0}-\e{p}{+}}{E^{+}_{\Delta}(p)} \Lambda^{+}_{p}
\right) \gamma^{\nu }\gamma^{0} \left(
\frac{p_{0}-q_{0}+\e{p-q}{-}}{E^{-}_{\Delta}(p-q)}
\Lambda^{-}_{p-q}
+\frac{p_{0}-q_{0}-\e{p-q}{+}}{E^{+}_{\Delta}(p-q)}
\Lambda^{+}_{p-q}
\right) \nonumber \\
&-& \gamma^{\mu} \gamma^{0} \left(
\frac{p_{0}+\e{p}{+}}{E^{+}_{\Delta}(p)} \Lambda^{-}_{p}
+\frac{p_{0}-\e{p}{-}}{E^{-}_{\Delta}(p)} \Lambda^{+}_{p}
\right) \gamma^{\nu} \gamma^{0} \left(
\frac{p_{0}-q_{0}+\e{p-q}{+}}{E^{+}_{\Delta}(p-q)}
\Lambda^{-}_{p-q}
+\frac{p_{0}-q_{0}-\e{p-q}{-}}{E^{-}_{\Delta}(p-q)}
\Lambda^{+}_{p-q}
\right)
\nonumber \\
&-& 2\gamma^{\mu} \gamma^{0} \left(
\frac{1}{p_{0}-\e{p}{-}} \Lambda^{-}_{p}
+\frac{1}{p_{0}+\e{p}{+}} \Lambda^{+}_{p} \right)
\gamma^{\nu} \gamma^{0}  \left(
\frac{1}{p_{0}-q_{0}-\e{p-q}{-}} \Lambda^{-}_{p-q}
+\frac{1}{p_{0}-q_{0}+\e{p-q}{+}} \Lambda^{+}_{p-q}
\right) \nonumber \\
&-& 2\gamma^{\mu} \gamma^{0} \left(
\frac{1}{p_{0}-\e{p}{+}} \Lambda^{-}_{p}
+\frac{1}{p_{0}+\e{p}{-}} \Lambda^{+}_{p} \right)
\gamma^{\nu} \gamma^{0}  \left(
\frac{1}{p_{0}-q_{0}-\e{p-q}{+}} \Lambda^{-}_{p-q}
+\frac{1}{p_{0}-q_{0}+\e{p-q}{-}} \Lambda^{+}_{p-q}
\right)
\Bigg].
\label{Pi-8}
\ea
Here we made use of the the following results for color-flavor traces:
\ba
\mbox{tr}_{cf}\left( {\cal I}_{1} T^{A} {\cal I}_{1} T^{B} \right)
&=& \left\{ \begin{array}{ll}
\delta^{AB} & A,B =1,2,3 ;\\ 
0 & A,B =4,5,6,7 ;\\
\frac{1}{3} & A,B =8 ;
\end{array}\right. \\ 
\mbox{tr}_{cf}\left(  {\cal I}_{2} T^{A} {\cal I}_{2} T^{B} \right)
&=&\left\{ \begin{array}{ll}
0 & A,B =1,\ldots,7 ;\\
\frac{2}{3} & A,B =8 ; 
\end{array}\right. \\ 
\mbox{tr}_{cf}\left(  {\cal I}_{1} T^{A} {\cal I}_{2} T^{B} \right)
&=& \left\{ \begin{array}{ll}
0 & A,B =1,2,3 ;\\
\frac{1}{2} \delta^{AB}+\frac{i}{4} \left(
\delta^{A4}\delta^{B5} -\delta^{A5}\delta^{B4}
+\delta^{A6}\delta^{B7}-\delta^{A7}\delta^{B6}
\right) & A,B =4,5,6,7 ; \\
0 & A,B =8 ;
\end{array}\right. \\
\mbox{tr}_{cf}\left[ \hat{\varepsilon} T^{A} \hat{\varepsilon}
\left( T^{B} \right)^{T}
\right] &=& \left\{ \begin{array}{ll}
-\delta^{AB} & A,B =1,2,3 ;\\
0  & A,B =4,5,6,7 ;\\
\frac{1}{3} & A,B =8.
\end{array}\right.
\ea

\section{Calculation of integrals}
\label{integrals}

In this Appendix, we calculate different type of integrals that appear in
the expression for the polarization tensor (see the previous Appendix).
For our purposes it is sufficient to consider the gluon momenta much less
than $\mu$, neglecting all corrections of order $q^{2}/ \mu^{2}$.  In this
approximation, the Dirac traces, containing the on-shell projection
operators, read
\ba
\mbox{tr}\left[\gamma^{\mu} \Lambda^{(\pm)}_{p}
\gamma^{\nu} \Lambda^{(\pm)}_{p-q}\right] &\simeq&
2\left( g^{\mu\nu} - g^{\mu 0}g^{\nu 0}
+ \frac{\vec{p}^{\mu}\vec{p}^{\nu}}{|\vec{p}|^{2}}
\right),
\label{tr1++}\\
\mbox{tr}\left[\gamma^{\mu} \Lambda^{(\pm)}_{p}
\gamma^{\nu} \Lambda^{(\mp)}_{p-q}\right] &\simeq&
2 \left( g^{\mu 0}g^{\nu 0}
-\frac{\vec{p}^{\mu}\vec{p}^{\nu}}{|\vec{p}|^{2}}
\right),
\label{tr1+-}\\
\mbox{tr}\left[\gamma^{\mu} \gamma^0 \Lambda^{(\pm)}_{p}
\gamma^{\nu} \gamma^0 \Lambda^{(\pm)}_{p-q}\right] &\simeq&
2\left(g^{\mu 0} \mp \frac{\vec{p}^{\mu}}{|\vec{p}|}\right)
\left(g^{\nu 0} \mp \frac{\vec{p}^{\nu}}{|\vec{p}|}\right),
\label{tr2++} \\
\mbox{tr}\left[\gamma^{\mu} \gamma^0 \Lambda^{(\pm)}_{p}
\gamma^{\nu} \gamma^0 \Lambda^{(\mp)}_{p-q}\right] &\simeq&
-2\left(g^{\mu\nu}-g^{\mu0}g^{\nu0}
+\frac{\vec{p}^{\mu}\vec{p}^{\nu}}{|\vec{p}|^{2}} 
\right) .
\label{tr2+-}
\ea
Also notice that in order to perform the integrations properly,
one should make the following replacements
in the denominators of propagators:
\ba
E^{\pm}_{\Delta} &\to& E^{\pm}_{\Delta} +i \varepsilon, \\
p_0 \pm \e{p}{-} &\to& p_0 \pm \e{p}{-}
\mp i \varepsilon \mbox{~sign}(\e{p}{-}), \\
p_0 \pm \e{p}{+} &\to& p_0 \pm \e{p}{+}
\mp i \varepsilon.
\ea
Form the definition in the previous Appendix, we see that 
the general structure of the polarization tensor reads
\begin{mathletters}
\ba
\Pi_{1}^{\mu\nu}(q) &=& 4\pi \alpha_{s} \left(
J_{\Delta}^{\mu\nu} (q)+I_{\Delta}^{\mu\nu}(q)
\right) , 
\label{B8a} \\
\Pi_{4}^{\mu\nu}(q) &=& 4\pi \alpha_{s} 
\tilde{I}_{\Delta}^{\mu\nu}(q) , 
\label{B8b} \\
\Pi_{8}^{\mu\nu}(q) &=& 
-\frac{4\pi \alpha_{s}}{3} \left(
J_{\Delta}^{\mu\nu} (q)-I_{\Delta}^{\mu\nu}(q)\right)
+\frac{8\pi \alpha_{s}}{3} I_{HDL}^{\mu\nu}(q). 
\label{B8c}
\ea
\label{B8}
\end{mathletters}
[Here we took into account the fact that the off-diagonal term in $\left.
\Pi^{AB,\mu\nu}(q)\right|_{A,B=4,5,6,7}$ is zero. To see this, one has to
calculate the corresponding integral in Eq.~(\ref{Pi-4567}).] Let us start
with the calculation of the first type of the integrals,
\ba
I_{HDL}^{\mu\nu}(q)&=&i\int\frac{d^{4} p}{(2\pi)^{4}} \mbox{tr}_{D}
\left[\gamma^{\mu} \gamma^{0} \left(
\frac{1}{p_{0}-\e{p}{-}} \Lambda^{-}_{p}
+\frac{1}{p_{0}+\e{p}{+}} \Lambda^{+}_{p} \right)
\gamma^{\nu} \gamma^{0}  \left(
\frac{1}{p_{0}-q_{0}-\e{p-q}{-}} \Lambda^{-}_{p-q}
+\frac{1}{p_{0}-q_{0}+\e{p-q}{+}} \Lambda^{+}_{p-q}
\right) \right] \nonumber \\
&=& i\int\frac{d^{4} p}{(2\pi)^{4}} \mbox{tr}_{D}
\left[\gamma^{\mu} \gamma^{0} \left(
\frac{1}{p_{0}-\e{p}{+}} \Lambda^{-}_{p}
+\frac{1}{p_{0}+\e{p}{-}} \Lambda^{+}_{p} \right)
\gamma^{\nu} \gamma^{0}  \left(
\frac{1}{p_{0}-q_{0}-\e{p-q}{+}} \Lambda^{-}_{p-q}
+\frac{1}{p_{0}-q_{0}+\e{p-q}{-}} \Lambda^{+}_{p-q}
\right)
\right] \nonumber \\
&=& -2\int \frac{d^{3}p}{(2\pi)^{3}} 
\left(g^{\mu 0}+\frac{\vec{p}^{\mu}}{|\vec{p}|}\right)
\left(g^{\nu 0}+\frac{\vec{p}^{\nu}}{|\vec{p}|}\right) 
\left( \frac{\theta(-\e{p}{-}) \theta(\e{p-q}{-}) }
{\e{p}{-}-\e{p-q}{-}-q_{0}+i\varepsilon} +
\frac{\theta(\e{p}{-}) \theta(-\e{p-q}{-})
}{\e{p-q}{-}-\e{p}{-}+q_{0}+i\varepsilon}
\right)\nonumber \\
&& -2 \int \frac{d^{3}p}{(2\pi)^{3}}
\left(g^{\mu\nu}-g^{\mu0}g^{\nu0}
+\frac{\vec{p}^{\mu}\vec{p}^{\nu}}{|\vec{p}|^{2}}
\right)\left(\frac{\theta(\e{p}{-})}{p}-\frac{1}{p}\right)
\nonumber \\
&=&-\frac{\mu^{2}}{\pi^{2}}\Bigg[g^{\mu0}g^{\nu0} 
Q\left(\frac{q_{0}}{q}\right)
-\frac{1}{2}\left(g^{\mu\nu}-g^{\mu0}g^{\nu0}
+\frac{\vec{q}^{\mu}\vec{q}^{\nu}}{q^{2}}
\right) \left(1+\frac{q^{2}-q_{0}^{2}}{q^{2}}
Q\left(\frac{q_{0}}{q}\right) \right) \nonumber \\
&& +\frac{\vec{q}^{\mu}\vec{q}^{\nu}}{q^{2}} 
\frac{q_{0}^{2}}{q^{2}} Q\left(\frac{q_{0}}{q}\right) 
+\frac{q_{0}}{q}
\left(g^{\mu 0}\frac{\vec{q}^{\nu}}{q}
+g^{\nu 0}\frac{\vec{q}^{\mu}}{q}\right)
Q\left(\frac{q_{0}}{q}\right) \Bigg] ,
\ea
where 
\be
Q\left(x\right)\equiv -\frac{1}{2} \int_{0}^{1} d \xi
\left(\frac{\xi}{\xi +x -i \varepsilon}+
\frac{\xi}{\xi -x -i \varepsilon}\right)
= \frac{x}{2} \ln \left|\frac{1+x}{1-x}\right|
-1 -i\frac{\pi}{2}|x| \theta(1-x^2).
\ee
It is easy to check that $q_{\mu}I^{\mu\nu}_{HDL}(q)=0$, as it should be.
Also notice that in Eucledian space, $x=iq_{4}/q \equiv iy$, and
\be
Q\left(x\right) \to \tilde{Q}\left(y\right)= 
-1+y \arctan\frac{1}{y} \simeq \left\{ \begin{array}{lll}
-1+\frac{\pi}{2} y , & \mbox{for} & y \ll 1;\\
-\frac{1}{3y^{2}}  , & \mbox{for} & y \gg 1.
\end{array}
\right.
\ee
Now, let us consider
\ba
J_{\Delta}^{\mu\nu}(q)&=&
i\int\frac{d^{4} p}{(2\pi)^{4}} \mbox{tr}_{D} 
\left[ \gamma^{\mu} \left(
\frac{\Delta^{+}\Lambda^{-}_{p}}{E^{+}_{\Delta}(p)}
+\frac{\Delta^{-}\Lambda^{+}_{p}}{E^{-}_{\Delta}(p)}
\right) \gamma^{\nu} \left(
\frac{(\Delta^{-})^{*}\Lambda^{-}_{p-q}}{E^{-}_{\Delta}(p-q)}
+\frac{(\Delta^{+})^{*}\Lambda^{+}_{p-q}}{E^{+}_{\Delta}(p-q)}
\right) \right] \nonumber \\
&=& i\int\frac{d^{4} p}{(2\pi)^{4}} \mbox{tr}_{D}
\left[ \gamma^{\mu} \left(
\frac{(\Delta^{-})^{*}\Lambda^{-}_{p}}{E^{-}_{\Delta}(p)}
+\frac{(\Delta^{+})^{*}\Lambda^{+}_{p}}{E^{+}_{\Delta}(p)}
\right) \gamma^{\nu} \left(
\frac{\Delta^{+}\Lambda^{-}_{p-q}}{E^{+}_{\Delta}(p-q)}
+\frac{\Delta^{-}\Lambda^{+}_{p-q}}{E^{-}_{\Delta}(p-q)}
\right)\right]\nonumber \\
&\simeq& 2 i \int \frac{d^{4} p}{(2\pi)^{4}} 
\left( g^{\mu 0}g^{\nu 0}
-\frac{\vec{p}^{\mu}\vec{p}^{\nu}}{p^{2}} \right)
\frac{|\Delta^{-}|^{2}}{E^{-}_{\Delta}(p) E^{-}_{\Delta}(p-q)}
\nonumber \\
&=&-\frac{\mu^2}{4\pi^{2}}
\int_{0}^{1} d \xi \int_{0}^{1} d x
\frac{|\Delta^{-}|^{2}\left[ (1+\xi^{2}) g^{\mu 0}g^{\nu 0}
+(1-\xi^{2}) g^{\mu\nu} +(1-3\xi^{2})
\frac{\vec{q}^{\mu}\vec{q}^{\nu}}{q^{2}} \right]}
{|\Delta^{-}|^{2}+x(1-x)\left(\xi^{2} q^{2} -q_{0}^{2}
\right)-i\varepsilon } .
\label{I-D-Min}
\ea
After switching to the Eucledian momenta ($q_{0}=iq_{4}$), we
get
\ba
J_{\Delta}^{\mu\nu}(q) &=&
-\frac{\mu^2}{4\pi^{2}}\int_{0}^{1} d \xi
\int_{0}^{1} d x
\frac{|\Delta^{-}|^{2}\left[ (1+\xi^{2}) g^{\mu 0}g^{\nu 0}
+(1-\xi^{2}) g^{\mu\nu} +(1-3\xi^{2})   
\frac{\vec{q}^{\mu}\vec{q}^{\nu}}{q^{2}}
\right]}
{|\Delta^{-}|^{2}+x(1-x)\left(q_{4}^{2} +\xi^{2} q^{2}\right)}
\nonumber \\
&=& - \frac{\mu^2}{2\pi^{2}}\int_{0}^{1} d \xi
\frac{|\Delta^{-}|^{2}\left[ (1+\xi^{2}) g^{\mu 0}g^{\nu 0}
+(1-\xi^{2}) g^{\mu\nu} +(1-3\xi^{2})
\frac{\vec{q}^{\mu}\vec{q}^{\nu}}{q^{2}}
\right]}{\sqrt{q_{4}^{2}+\xi^{2} q^{2}}
\sqrt{q_{4}^{2}+\xi^{2}q^{2}+4|\Delta^{-}|^{2}} } \nonumber \\
&&\times
\ln\frac{\sqrt{q_{4}^{2}+\xi^{2}q^{2}+4|\Delta^{-}|^{2}}+
\sqrt{q_{4}^{2}+\xi^{2}q^{2}}}
{\sqrt{q_{4}^{2}+\xi^{2}q^{2}+4|\Delta^{-}|^{2}}-   
\sqrt{q_{4}^{2}+\xi^{2}q^{2}}} .
\label{I-D-Euc}
\ea
In different limits, we obtain the following behavior:
\ba
J_{\Delta}^{\mu\nu}(q)= \left\{ \begin{array}{lll}
-\frac{\mu^2}{3\pi^{2}}
\left( g^{\mu 0}g^{\nu 0} +\frac{1}{2} g^{\mu\nu} \right)
+\delta J^{(2)}(q),
& \mbox{for} & |q_{4}|,q \ll 2|\Delta^{-}|; \\
-\frac{2\mu^2 |\Delta^{-}|^{2}}{3\pi^{2} q_{4}^{2}}
\ln\frac{q_{4}^{2}}{|\Delta^{-}|^{2}}
\left( g^{\mu 0}g^{\nu 0} + \frac{1}{2}g^{\mu\nu}\right),
& \mbox{for} & q, |\Delta^{-}| \ll |q_{4}|; \\
-\frac{\mu^2 |\Delta^{-}|^{2}}{4\pi |q_{4}| q}
\ln\frac{4q_{4}^{2}}{|\Delta^{-}|^{2}}
\left(g^{\mu 0}g^{\nu 0}+g^{\mu\nu}
+\frac{\vec{q}^{\mu}\vec{q}^{\nu}}{q^{2}}
\right) ,
& \mbox{for} & |\Delta^{-}| \ll |q_{4}| \ll q;\\
-\frac{\mu^2 |\Delta^{-}|}{8 q}
\left(g^{\mu 0}g^{\nu 0}+g^{\mu\nu}
+\frac{\vec{q}^{\mu}\vec{q}^{\nu}}{q^{2}}
\right) ,
& \mbox{for} &  |q_{4}| \ll |\Delta^{-}| \ll q ,
\end{array}
\right.
\label{B16}
\ea
where (in Minkowski space)
\be
\delta J^{(2)}(q) = \frac{\mu^2}{90\pi^{2}|\Delta^{-}|^{2}} \left[
g^{\mu 0}g^{\nu 0} (2q^{2}-5q_{0}^{2})   
+\frac{1}{2}g^{\mu\nu} (q^{2}-5q_{0}^{2})
- \vec{q}^{\mu}\vec{q}^{\nu} \right].
\label{del-J-2}
\ee

As one could see, the imaginary part of $J_{\Delta}^{\mu\nu}(q)$ 
in Eq.~(\ref{I-D-Min}) appears only for $q_0 > 2
|\Delta^{-}|$. Its explicit form is given in terms of the 
elliptic integral of the first, $F(\varphi,z)$, and second kind, 
$E(\varphi,z)$, and the hypergeometric function, 
\ba
&&\mbox{Im} \left[ J_{\Delta}^{\mu\nu}(q) \right] 
=\frac{\mu^{2}|\Delta^{-}|^{2}}{2\pi} 
\frac{\theta\left(q_{0}^{2}-q^{2}-4|\Delta^{-}|^{2}\right)}
{\sqrt{q^{2}} \sqrt{q_{0}^{2}-4|\Delta^{-}|^{2}}} 
\Bigg\{\left(g^{\mu\nu} +g^{\mu 0}g^{\nu 0}
+\frac{\vec{q}^{\mu}\vec{q}^{\nu}}{q^{2}}\right)
F\left(\arcsin\sqrt{\frac{q^{2}}{q_{0}^{2}}},
\frac{q_{0}^{2}}{q_{0}^{2}-4|\Delta^{-}|^{2}}\right)
\nonumber \\
&&-\frac{q_{0}^{2}-4|\Delta^{-}|^{2}}{q_{0}^{2}}
\left(g^{\mu\nu} -g^{\mu 0}g^{\nu 0}
+3\frac{\vec{q}^{\mu}\vec{q}^{\nu}}{q^{2}}\right)
\left[F\left(\arcsin\sqrt{\frac{q^{2}}{q_{0}^{2}}},
\frac{q_{0}^{2}}{q_{0}^{2}-4|\Delta^{-}|^{2}}\right)
-E\left(\arcsin\sqrt{\frac{q^{2}}{q_{0}^{2}}},
\frac{q_{0}^{2}}{q_{0}^{2}-4|\Delta^{-}|^{2}}\right)
\right]\Bigg\} \nonumber \\
&+&\frac{\mu^{2}|\Delta^{-}|^{2}}{2\pi} 
\frac{\theta\left(4|\Delta^{-}|^{2}+q^{2}-q_{0}^{2}\right)
\theta\left(q_{0}^{2}-4|\Delta^{-}|^{2}\right)}
{\sqrt{q^{2}} \sqrt{q_{0}^{2}}}
\Bigg\{\left(g^{\mu\nu} +g^{\mu 0}g^{\nu 0}
+\frac{\vec{q}^{\mu}\vec{q}^{\nu}}{q^{2}}\right)
K\left(\frac{q_{0}^{2}-4|\Delta^{-}|^{2}}{q_{0}^{2}}
\right)\nonumber \\
&&-\frac{\pi(q_{0}^{2}-4|\Delta^{-}|^{2})}{4q^{2}} 
\left(g^{\mu\nu} -g^{\mu 0}g^{\nu 0}
+3\frac{\vec{q}^{\mu}\vec{q}^{\nu}}{q^{2}}\right)
~_{2}F_{1}\left(\frac{1}{2},\frac{3}{2},2,
\frac{q_{0}^{2}-4|\Delta^{-}|^{2}}{q_{0}^{2}}\right)
\Bigg\} .
\ea
Similarly, let us calculate the following quantity:
\ba
I_{\Delta}^{\mu\nu}(q)&=&
i\int\frac{d^{4} p}{(2\pi)^{4}} \mbox{tr}_{D}
\left[\gamma^{\mu} \gamma^{0} \left(
\frac{p_{0}+\e{p}{-}}{E^{-}_{\Delta}(p)} \Lambda^{-}_{p} 
+\frac{p_{0}-\e{p}{+}}{E^{+}_{\Delta}(p)} \Lambda^{+}_{p}
\right) \gamma^{\nu }\gamma^{0} \left(
\frac{p_{0}-q_{0}+\e{p-q}{-}}{E^{-}_{\Delta}(p-q)}
\Lambda^{-}_{p-q}
+\frac{p_{0}-q_{0}-\e{p-q}{+}}{E^{+}_{\Delta}(p-q)}
\Lambda^{+}_{p-q}
\right) \right] \nonumber \\
&=& i\int\frac{d^{4} p}{(2\pi)^{4}} \mbox{tr}_{D}
\left[ \gamma^{\mu} \gamma^{0} \left(
\frac{p_{0}+\e{p}{+}}{E^{+}_{\Delta}(p)} \Lambda^{-}_{p}
+\frac{p_{0}-\e{p}{-}}{E^{-}_{\Delta}(p)} \Lambda^{+}_{p}
\right) \gamma^{\nu} \gamma^{0} \left(
\frac{p_{0}-q_{0}+\e{p-q}{+}}{E^{+}_{\Delta}(p-q)}
\Lambda^{-}_{p-q}
+\frac{p_{0}-q_{0}-\e{p-q}{-}}{E^{-}_{\Delta}(p-q)}
\Lambda^{+}_{p-q}
\right) \right] \nonumber \\  
&\simeq& \frac{\mu^{2}}{3\pi^{2}}
\left(g^{\mu\nu}-g^{\mu 0}g^{\nu 0}\right)
+2i\int\frac{d^{4} p}{(2\pi)^{4}} 
\left(g^{\mu 0} + \frac{\vec{p}^{\mu}}{|\vec{p}|}\right)
\left(g^{\nu 0} + \frac{\vec{p}^{\nu}}{|\vec{p}|}\right)
\frac{(p_{0}+\e{p}{-})(p_{0}-q_{0}+\e{p-q}{-})}
{E^{-}_{\Delta}(p) E^{-}_{\Delta}(p-q)} \nonumber \\  
&\simeq& \frac{\mu^{2}}{3\pi^{2}}
\left(g^{\mu\nu}-g^{\mu 0}g^{\nu 0}\right)
+2i \int_{0}^{1} dx \int \frac{d p_{0} d^{3} p}{(2\pi)^{4}}
\left(g^{\mu 0} + \frac{\vec{p}^{\mu}}{|\vec{p}|}\right)
\left(g^{\nu 0} + \frac{\vec{p}^{\nu}}{|\vec{p}|}\right)
\nonumber \\
&\times&
\frac{\left[p_{0}+\e{p}{-}+x(q_{0}+\xi q)\right] 
\left[p_{0}+\e{p}{-}-(1-x)(q_{0}+\xi q)\right] }
{\left[p_{0}^{2}-(\e{p}{-})^{2}
-x(1-x)(\xi^{2}q^{2}-q_{0}^{2}) -|\Delta^{-}|^{2}
+i\varepsilon\right]^{2}} .
\ea
So, we derive
\ba
I_{\Delta}^{\mu\nu}(q) 
&=& \frac{\mu^{2}}{3\pi^{2}}
\left(g^{\mu\nu}-g^{\mu 0}g^{\nu 0}\right)
+\frac{\mu^{2}}{4\pi^{2}} \int_{0}^{1} dx 
\int_{-1}^{1} d\xi 
\frac{|\Delta^{-}|^{2}+2x(1-x)(q_{0} q \xi +q^{2} \xi^{2})}
{|\Delta^{-}|^{2} + x(1-x)(q^{2} \xi^{2}-q_{0}^{2})
-i\varepsilon} \nonumber \\
& &\times \left[
\frac{1}{2}(3-\xi^{2}) g^{\mu 0} g^{\nu 0} 
- \frac{1}{2}(1-\xi^{2}) g^{\mu\nu}
-\frac{1}{2}(1-3\xi^{2}) 
\frac{\vec{q}^{\mu}\vec{q}^{\nu}}{q^{2}} 
+\xi \left(g^{\mu 0}\frac{\vec{q}^{\nu}}{q}
+g^{\nu 0}\frac{\vec{q}^{\mu}}{q}\right) \right] 
\nonumber \\
&=&\frac{\mu^{2}}{3\pi^{2}}
\left(g^{\mu\nu}-g^{\mu 0}g^{\nu 0}\right)
+\frac{\mu^{2}}{4\pi^{2}} \int_{0}^{1} dx 
\int_{0}^{1} d\xi \left[
(3-\xi^{2}) g^{\mu 0} g^{\nu 0} 
- (1-\xi^{2}) g^{\mu\nu}
-(1-3\xi^{2}) 
\frac{\vec{q}^{\mu}\vec{q}^{\nu}}{q^{2}} 
\right] \nonumber \\
& &\times 
\frac{|\Delta^{-}|^{2}+2x(1-x) q^{2} \xi^{2}}
{|\Delta^{-}|^{2} + x(1-x)(q^{2} \xi^{2}-q_{0}^{2})
-i\varepsilon} \nonumber \\
&+& \frac{\mu^{2}}{\pi^{2}} \int_{0}^{1} dx 
\int_{0}^{1} d\xi 
\left(g^{\mu 0}\frac{\vec{q}^{\nu}}{q}
+g^{\nu 0}\frac{\vec{q}^{\mu}}{q}\right)
\frac{x(1-x) q q_{0} \xi^{2}}
{|\Delta^{-}|^{2} + x(1-x)(q^{2} \xi^{2}-q_{0}^{2})
-i\varepsilon} .
\ea
After switching to Eucledian momenta ($q_{0} = i q_{4}$), we could 
also perform the integration over $x$, and get the following 
result:
\ba
I_{\Delta}^{\mu\nu}(q) &=& I_{HDL}^{\mu\nu}(q) 
+\frac{\mu^{2}}{2\pi^{2}} \int_{0}^{1} d\xi \left[
(3-\xi^{2}) g^{\mu 0} g^{\nu 0}
- (1-\xi^{2}) g^{\mu\nu}
-(1-3\xi^{2})
\frac{\vec{q}^{\mu}\vec{q}^{\nu}}{q^{2}}
\right] \nonumber \\
& &\times 
\frac{|\Delta^{-}|^{2}(q_{4}^{2}-\xi^{2}q^{2})}
{\sqrt{q_{4}^{2}+\xi^{2}q^{2}+4|\Delta^{-}|^{2}}
(q_{4}^{2}+\xi^{2}q^{2})^{3/2}}
\ln\frac{\sqrt{q_{4}^{2}+\xi^{2}q^{2}+4|\Delta^{-}|^{2}}+
\sqrt{q_{4}^{2}+\xi^{2}q^{2}}}
{\sqrt{q_{4}^{2}+\xi^{2}q^{2}+4|\Delta^{-}|^{2}}-
\sqrt{q_{4}^{2}+\xi^{2}q^{2}}} \nonumber \\
&-& \frac{2\mu^{2}}{\pi^{2}} \frac{iq_{4}}{q}
\int_{0}^{1} d\xi
\frac{\left(g^{\mu 0}\frac{\vec{q}^{\nu}}{q}
+g^{\nu 0}\frac{\vec{q}^{\mu}}{q}\right) 
|\Delta^{-}|^{2} \xi^{2} q^{2}}
{\sqrt{q_{4}^{2}+\xi^{2}q^{2}+4|\Delta^{-}|^{2}}
(q_{4}^{2}+\xi^{2}q^{2})^{3/2}}
\ln\frac{\sqrt{q_{4}^{2}+\xi^{2}q^{2}+4|\Delta^{-}|^{2}}+
\sqrt{q_{4}^{2}+\xi^{2}q^{2}}}
{\sqrt{q_{4}^{2}+\xi^{2}q^{2}+4|\Delta^{-}|^{2}}-
\sqrt{q_{4}^{2}+\xi^{2}q^{2}}} . 
\ea
The following asymptotes are valid
\ba
I_{\Delta}^{\mu\nu}(q) -I_{HDL}(q)
= \left\{ \begin{array}{lll}
-I_{HDL}(q)+\frac{\mu^2 }{3 \pi^{2}}
\left(g^{\mu 0}g^{\nu 0} +\frac{1}{2}g^{\mu\nu}\right)
+\delta I^{(2)}(q),
& \mbox{for} & |q_{4}|,q \ll 2|\Delta^{-}|; \\
\frac{\mu^2 |\Delta^{-}|^{2}}{3\pi^{2} q_{4}^{2}}
\ln\frac{q_{4}^{2}}{|\Delta^{-}|^{2}}
\left( 4 g^{\mu 0}g^{\nu 0} - g^{\mu\nu} 
+2\frac{g^{\mu 0}\vec{q}^{\nu}+\vec{q}^{\mu}g^{\nu 0}}
{iq_{4}} \right),
& \mbox{for} & q, |\Delta^{-}| \ll |q_{4}|; \\
-\frac{\mu^2 |\Delta^{-}|^{2}}{4\pi |q_{4}| q}
\ln\frac{4q_{4}^{2}}{|\Delta^{-}|^{2}}
\left(3g^{\mu 0}g^{\nu 0}-g^{\mu\nu}
-\frac{\vec{q}^{\mu}\vec{q}^{\nu}}{q^{2}}
+4iq_{4}\frac{g^{\mu 0}\vec{q}^{\nu}+\vec{q}^{\mu}g^{\nu 0}}
{q^{2}} \right) ,
& \mbox{for} & |\Delta^{-}| \ll |q_{4}| \ll q;\\
-\frac{\mu^2 |\Delta^{-}|}{8 q}
\left(3g^{\mu 0}g^{\nu 0}-g^{\mu\nu}
-\frac{\vec{q}^{\mu}\vec{q}^{\nu}}{q^{2}}
+4iq_{4}\frac{g^{\mu 0}\vec{q}^{\nu}+\vec{q}^{\mu}g^{\nu 0}}
{q^{2}} \right) ,
& \mbox{for} &  |q_{4}| \ll |\Delta^{-}| \ll q.
\end{array}
\right.
\label{B21}
\ea
where (in Minkowski space) $\delta I^{(2)}(q)$ reads
\be
\delta I^{(2)}(q) = \frac{\mu^2}{90\pi^{2}|\Delta^{-}|^{2}} \left[
g^{\mu 0}g^{\nu 0} (3q^{2}+10 q_{0}^{2})
-\frac{1}{2}g^{\mu\nu} (q^{2}+5 q_{0}^{2})
+ \vec{q}^{\mu}\vec{q}^{\nu} \right]
+\frac{\mu^2 q_{0}
\left( \vec{q}^{\mu} u^{\nu} + u^{\mu} \vec{q}^{\nu} \right)}
{18\pi^{2}|\Delta^{-}|^{2}} . 
\label{del-I-2}
\ee
Finally, we calculate the last type of integrals,
\ba
\tilde{I}_{\Delta}^{\mu\nu}(q)&=&
i\int\frac{d^{4} p}{(2\pi)^{4}} \mbox{tr}_{D}
\left[\gamma^{\mu} \gamma^{0} \left(
\frac{p_{0}+\e{p}{-}}{E^{-}_{\Delta}(p)} \Lambda^{-}_{p} 
+\frac{p_{0}-\e{p}{+}}{E^{+}_{\Delta}(p)} \Lambda^{+}_{p}
\right) \gamma^{\nu }\gamma^{0} \left(
\frac{1}{p_{0}-q_{0}-\e{p-q}{-}} \Lambda^{-}_{p-q}
+\frac{1}{p_{0}-q_{0}+\e{p-q}{+}} \Lambda^{+}_{p-q}
\right) \right] \nonumber \\
&=& i\int\frac{d^{4} p}{(2\pi)^{4}} \mbox{tr}_{D}
\left[ \gamma^{\mu} \gamma^{0} \left(
\frac{p_{0}+\e{p}{+}}{E^{+}_{\Delta}(p)} \Lambda^{-}_{p}
+\frac{p_{0}-\e{p}{-}}{E^{-}_{\Delta}(p)} \Lambda^{+}_{p}
\right)\gamma^{\nu} \gamma^{0}  \left(
\frac{1}{p_{0}-q_{0}-\e{p-q}{+}} \Lambda^{-}_{p-q}
+\frac{1}{p_{0}-q_{0}+\e{p-q}{-}} \Lambda^{+}_{p-q}
\right)
\right] \nonumber \\
&\simeq& \frac{\mu^{2}}{3\pi^{2}}
\left(g^{\mu\nu}-g^{\mu 0}g^{\nu 0}\right)
+2i\int\frac{d^{4} p}{(2\pi)^{4}} 
\left(g^{\mu 0} + \frac{\vec{p}^{\mu}}{|\vec{p}|}\right)
\left(g^{\nu 0} + \frac{\vec{p}^{\nu}}{|\vec{p}|}\right)
\frac{(p_{0}+\e{p}{-})(p_{0}-q_{0}+\e{p-q}{-})}
{[p_{0}^{2}-(\e{p}{-})^{2}-|\Delta^{-}|^{2}] E^{-}_{\Delta}(p-q)} 
\nonumber \\  
&\simeq&\frac{\mu^{2}}{3\pi^{2}}
\left(g^{\mu\nu}-g^{\mu 0}g^{\nu 0}\right)
+2i \int_{0}^{1} dx \int \frac{d p_{0} d^{3} p}{(2\pi)^{4}}
\left(g^{\mu 0} + \frac{\vec{p}^{\mu}}{|\vec{p}|}\right)
\left(g^{\nu 0} + \frac{\vec{p}^{\nu}}{|\vec{p}|}\right)
\nonumber \\
&\times&
\frac{\left[p_{0}+\e{p}{-}+x(q_{0}+\xi q)\right] 
\left[p_{0}+\e{p}{-}-(1-x)(q_{0}+\xi q)\right] }
{\left[p_{0}^{2}-(\e{p}{-})^{2}
-x(1-x)(\xi^{2}q^{2}-q_{0}^{2}) -(1-x)|\Delta^{-}|^{2}
+i\varepsilon\right]^{2}}\nonumber \\
&\simeq&\frac{\mu^{2}}{3\pi^{2}}
\left(g^{\mu\nu}-g^{\mu 0}g^{\nu 0}\right)
+\frac{\mu^{2}}{4\pi^{2}} \int_{0}^{1} dx 
\int_{0}^{1} d\xi \frac{\left[
(3-\xi^{2}) g^{\mu 0} g^{\nu 0} 
- (1-\xi^{2}) g^{\mu\nu}
-(1-3\xi^{2}) 
\frac{\vec{q}^{\mu}\vec{q}^{\nu}}{q^{2}} 
\right] (|\Delta^{-}|^{2}+2x q^{2} \xi^{2}) }
{|\Delta^{-}|^{2} + x(q^{2} \xi^{2}-q_{0}^{2})
-i\varepsilon} \nonumber \\
&&+ \frac{\mu^{2}}{\pi^{2}} \int_{0}^{1} dx 
\int_{0}^{1} d\xi 
\left(g^{\mu 0}\frac{\vec{q}^{\nu}}{q}
+g^{\nu 0}\frac{\vec{q}^{\mu}}{q}\right)
\frac{x q q_{0} \xi^{2}}
{|\Delta^{-}|^{2} + x(q^{2} \xi^{2}-q_{0}^{2})
-i\varepsilon}.
\ea
After switching to Eucledian momenta ($q_{0} = i q_{4}$), we could 
also perform the integration over $x$, and get the following 
result:
\ba
\tilde{I}_{\Delta}^{\mu\nu}(q)
&=& I_{HDL}^{\mu\nu}(q)
+\frac{\mu^{2}}{4\pi^{2}} \int_{0}^{1} d\xi \left[
(3-\xi^{2}) g^{\mu 0} g^{\nu 0}
- (1-\xi^{2}) g^{\mu\nu}
-(1-3\xi^{2})
\frac{\vec{q}^{\mu}\vec{q}^{\nu}}{q^{2}}
\right] 
\frac{|\Delta^{-}|^{2}(q_{4}^{2}-\xi^{2}q^{2})}
{(q_{4}^{2}+\xi^{2}q^{2})^{2} }
\ln\frac{|\Delta^{-}|^{2}+q_{4}^{2}+\xi^{2}q^{2}}{|\Delta^{-}|^{2}}
\nonumber \\
&-& \frac{\mu^{2}}{\pi^{2}} \frac{iq_{4}}{q}
\int_{0}^{1} d\xi
\left(g^{\mu 0}\frac{\vec{q}^{\nu}}{q}
+g^{\nu 0}\frac{\vec{q}^{\mu}}{q}\right)
\frac{|\Delta^{-}|^{2} \xi^{2} q^{2}}
{(q_{4}^{2}+\xi^{2}q^{2})^{2} }
\ln\frac{|\Delta^{-}|^{2}+q_{4}^{2}+\xi^{2}q^{2}}
{|\Delta^{-}|^{2}}.
\ea
The following asymptotes are valid
\ba
\tilde{I}_{\Delta}^{\mu\nu}(q) -I_{HDL}(q)
= \left\{ \begin{array}{lll}
-I_{HDL}(q)+\frac{\mu^2 }{3 \pi^{2}}
\left(g^{\mu 0}g^{\nu 0} +\frac{1}{2}g^{\mu\nu}\right)
+\delta \tilde{I}^{(2)}(q),
& \mbox{for} & |q_{4}|,q \ll 2|\Delta^{-}|; \\
\frac{\mu^2 |\Delta^{-}|^{2}}{6\pi^{2} q_{4}^{2}}
\ln\frac{q_{4}^{2}}{|\Delta^{-}|^{2}}
\left( 4 g^{\mu 0}g^{\nu 0} - g^{\mu\nu} 
+2\frac{g^{\mu 0}\vec{q}^{\nu}+\vec{q}^{\mu}g^{\nu 0}}
{iq_{4}} \right),
& \mbox{for} & q, |\Delta^{-}| \ll |q_{4}|; \\
-\frac{\mu^2 |\Delta^{-}|^{2}}{8\pi |q_{4}| q}
\left(3g^{\mu 0}g^{\nu 0}-g^{\mu\nu}
-\frac{\vec{q}^{\mu}\vec{q}^{\nu}}{q^{2}}
+iq_{4}\frac{g^{\mu 0}\vec{q}^{\nu}+\vec{q}^{\mu}g^{\nu 0}}
{q^{2}} \right), 
& \mbox{for} & |\Delta^{-}| \ll |q_{4}| \ll q;\\
-\frac{\mu^2 |\Delta^{-}|}{4 \pi q}
\left(3g^{\mu 0}g^{\nu 0}-g^{\mu\nu}
-\frac{\vec{q}^{\mu}\vec{q}^{\nu}}{q^{2}}
+iq_{4}\frac{g^{\mu 0}\vec{q}^{\nu}+\vec{q}^{\mu}g^{\nu 0}}
{q^{2}} \right) ,
& \mbox{for} &  |q_{4}| \ll |\Delta^{-}| \ll q.
\end{array}\right.
\label{B24}
\ea
where (in Minkowski space)
\be
\delta \tilde{I}^{(2)}(q) 
= \frac{\mu^2}{30\pi^{2}|\Delta^{-}|^{2}} \left[
g^{\mu 0}g^{\nu 0} (3q^{2}+10q_{0}^{2})
-\frac{1}{2}g^{\mu\nu} (q^{2}+5q_{0}^{2})
+ \vec{q}^{\mu}\vec{q}^{\nu} \right]
+\frac{\mu^2 q_{0}
\left( \vec{q}^{\mu} g^{\nu 0} + g^{\mu 0} \vec{q}^{\nu} \right)}
{5\pi^{2}|\Delta^{-}|^{2}}.
\ee

\section{Gauge invariance and polarization tensor}
\label{tra-pol-ten}

By making use of the definitions in Eq.~(\ref{B8}) as well as the general
structure of the tensors $J_{\Delta}^{\mu\nu}(q)$,
$I_{\Delta}^{\mu\nu}(q)$ and $\tilde{I}_{\Delta}^{\mu\nu}(q)$, given in
the preceding Appendix, it is easy to show that all three polarization
tensors $\Pi_{i}^{\mu\nu}(q) $ ($i=1,4,8$) take the following general
form:
\be
\Pi_{i}^{\mu\nu}(q) = \Pi_{i}^{(1)}(q) O^{(1)\mu\nu}(q)
+\Pi_{i}^{(2)}(q) O^{(2)\mu\nu}(q)
+\Pi_{i}^{(3)}(q) O^{(3)\mu\nu}(q)
+\Pi_{i}^{(4)}(q) O^{(4)\mu\nu}(q) .
\label{Pi-i-decomp}
\ee
In this representation, we use the following set of four tensors
\cite{Gusynin:2001ib}:
\begin{mathletters}
\ba
O^{(1)}_{\mu\nu}(q)&=& g_{\mu\nu}-u_{\mu} u_{\nu}
+\frac{\vec{q}_{\mu}\vec{q}_{\nu}}{|\vec{q}|^{2}},
\label{def-O1} \\
O^{(2)}_{\mu\nu}(q)&=& u_{\mu} u_{\nu}
-\frac{\vec{q}_{\mu}\vec{q}_{\nu}}{|\vec{q}|^{2}}
-\frac{q_{\mu}q_{\nu}}{q^{2}},
\label{def-O2} \\
O^{(3)}_{\mu\nu}(q)&=& \frac{q_{\mu}q_{\nu}}{q^{2}},
\label{def-O3} \\
O^{(4)}_{\mu \nu}(q) &=&
O^{(2)}_{\mu \lambda} u^{\lambda} \frac{q_{\nu}}{|\vec{q}|}
+\frac{q_{\mu}}{|\vec{q}|}u^{\lambda} O^{(2)}_{\lambda \nu}.
\label{def-O4}
\ea
\label{def-Os}
\end{mathletters}
The first three of them are the same projectors of the magnetic, electric
and unphysical (longitudinal in a 3+1 dimensional sense) modes of gluons
which were used in Ref.~\cite{Hong:2000fh}. In addition, here we also
introduced the intervening operator $O^{(4)}_{\mu \nu}(q)$, mixing the
electric and unphysical modes \cite{Gusynin:2001ib}. Note that $u_{\mu} 
=(1,0,0,0)$ and $\vec{q}_{\mu} =q_{\mu} - (u\cdot q) u_{\mu}$.

By making use of the representations for different types of 
the polarization tensors in Appendix~\ref{integrals}, we derive the
following explicit infrared ($|q_{0}|,q \ll |\Delta^{-}|$) asymptotes:
\ba
\Pi_{1}^{\mu\nu}(q) &=& \frac{2 \alpha_{s} \mu^{2} }
{9 \pi |\Delta^{-}|^{2}}
\left[ g^{\mu 0}g^{\nu 0} (q^{2}+q_{0}^{2}) -q_{0}^{2} g^{\mu\nu}
+q_{0} \left( \vec{q}^{\mu} g^{\nu 0} + g^{\mu 0} \vec{q}^{\nu} \right)
\right] \nonumber \\
&=& - \frac{2 \alpha_{s} \mu^{2} }
{9 \pi |\Delta^{-}|^{2}}
\left[ q_{0}^{2} O^{(1)\mu\nu}(q)
+(q_{0}^{2}-q^{2}) O^{(2)\mu\nu}(q) \right] , 
\label{Pi_1-IR} \\
\Pi_{4}^{\mu\nu}(q) &=& \frac{3}{2} \Pi_{8}^{\mu\nu}(q)
 = \frac{ 2 \alpha_{s} \mu^{2} }{ 3\pi } \left[
O^{(1)\mu\nu}(q)
+ \frac{q_{0}^{2}-3q^{2}}{q_{0}^{2}-q^{2}} O^{(2)\mu\nu}(q)
+ \frac{3q_{0}^{2}-q^{2}}{q_{0}^{2}-q^{2}} O^{(3)\mu\nu}(q)
+ \frac{2q_{0}q}{q_{0}^{2}-q^{2}} O^{(4)\mu\nu}(q) \right] .
\label{Pi_48-IR}
\ea
Similarly, we derive the asymptotes in other regions of interest:
\be
\Pi_{1}^{\mu\nu}(q) - 4\pi\alpha_{s} I_{HDL}^{\mu\nu}(q)
=  \left\{ \begin{array}{lll}
-\frac{8\alpha_{s}\mu^{2}|\Delta^{-}|^{2}}{3\pi q_{4}^{2}}
\ln\frac{q_{4}^{2}}{|\Delta^{-}|^{2}}\left[O^{(1)}(q)+O^{(2)}(q)\right],
& \mbox{for} & q, |\Delta^{-}| \ll |q_{4}|; \\
-\frac{\alpha_{s}\mu^{2}|\Delta^{-}|^{2}}{|q_{4}| q}
\ln\frac{4q_{4}^{2}}{|\Delta^{-}|^{2}}O^{(1)}(q), 
& \mbox{for} & |\Delta^{-}| \ll |q_{4}| \ll q;\\
-\frac{2\pi \alpha_{s}\mu^{2}|\Delta^{-}|}{q} O^{(2)}(q) ,
& \mbox{for} &  |q_{4}| \ll |\Delta^{-}| \ll q,
\end{array}\right.
\label{Pi_1-general}
\ee
\be
\Pi_{4}^{\mu\nu}(q) - 4\pi\alpha_{s} I_{HDL}^{\mu\nu}(q)
=  \left\{ \begin{array}{lll}
-\frac{2 \alpha_{s} \mu^2 |\Delta^{-}|^{2}}{3 \pi q_{4}^{2}}
\ln\frac{q_{4}^{2}}{|\Delta^{-}|^{2}} \left[
O^{(1)}(q)+O^{(2)}(q) -3 O^{(3)}(q)+\frac{2q}{iq_{4}}O^{(4)}(q) 
\right],
& \mbox{for} & q, |\Delta^{-}| \ll |q_{4}|; \\
\frac{\alpha_{s} \mu^2 |\Delta^{-}|^{2}}{2 |q_{4}| q}
\left[O^{(1)}(q)-O^{(2)}(q) +\frac{iq_{4}}{q}O^{(4)}(q)\right], 
& \mbox{for} & |\Delta^{-}| \ll |q_{4}| \ll q;\\
\frac{\pi \alpha_{s} \mu^2 |\Delta^{-}|}{q}
\left[O^{(1)}(q)-O^{(2)}(q) +\frac{iq_{4}}{q}O^{(4)}(q)\right] ,
& \mbox{for} &  |q_{4}| \ll |\Delta^{-}| \ll q,
\end{array}\right.
\label{Pi_4-general}
\ee
and 
\be
\Pi_{8}^{\mu\nu}(q)- 4\pi\alpha_{s} I_{HDL}^{\mu\nu}(q)
=  \left\{ \begin{array}{lll}
\frac{8\alpha_{s} \mu^2 |\Delta^{-}|^{2}}{3\pi q_{4}^{2}}
\ln\frac{q_{4}^{2}}{|\Delta^{-}|^{2}}
\left[O^{(3)}(q)+\frac{2q}{3iq_{4}}O^{(4)}(q)\right],
& \mbox{for} & q, |\Delta^{-}| \ll |q_{4}|; \\
\frac{2\alpha_{s} \mu^2 |\Delta^{-}|^{2}}{3 |q_{4}| q}
\ln\frac{4q_{4}^{2}}{|\Delta^{-}|^{2}}
\left[O^{(1)}(q)+\frac{4iq_{4}}{q} O^{(4)}(q)\right],
& \mbox{for} & |\Delta^{-}| \ll |q_{4}| \ll q;\\
\frac{\pi \alpha_{s} \mu^2 |\Delta^{-}|}{3q}
\left[O^{(1)}(q)+\frac{4iq_{4}}{q} O^{(4)}(q)\right] ,
& \mbox{for} &  |q_{4}| \ll |\Delta^{-}| \ll q.
\end{array}\right.
\label{Pi_8-general}
\ee
As is easy to check, $\Pi_{1}^{\mu\nu}(q)$ is transverse everywhere, while
$\Pi_{4}^{\mu\nu}(q)$ and $\Pi_{8}^{\mu\nu}(q)$ are not. From the
structure of the polarization tensor $\Pi_{1}^{\mu\nu}(q)$ in
Eq.~(\ref{Pi_1-IR}), we derive the explicit expression for the dielectric
constant in the far infrared region where the three gluons of the unbroken
$SU(2)_{c}$ decouple from the other degrees of freedom.  It coincides with
the result in Ref.~\cite{Rischke:2000cn} which is quoted in
Eq.~(\ref{diel-constant}). We also see that the magnetic permeability is
equal 1, because there is no magnetic contribution $q^{2}
O^{(1)\mu\nu}(q)$ on the right hand side of Eq.~(\ref{Pi_1-IR}) [notice
that $q_{0}^{2} O^{(1)\mu\nu}(q)$ is the electric type contribution]. Let
us also notice that the general expression for $\Pi_{1}^{\mu\nu}(q)$ is
transverse for all momenta. To see this, we derive the following
representation:
\ba
\Pi_{1}^{\mu\nu}(q) &=& 4\pi\alpha_{s} I_{HDL}^{\mu\nu}(q)
-\frac{4\alpha_{s} \mu^{2} |\Delta^{-}|^{2}}{\pi} \int_{0}^{1} 
\frac{ d \xi \left[ 
q_{4}^{2}(1-\xi^{2}) O^{(1)\mu\nu}(q)
+2 \xi^{2} (q_{4}^{2}+q^{2}) O^{(2)\mu\nu}(q) \right]}
{\sqrt{q_{4}^{2}+\xi^{2}q^{2}+4|\Delta^{-}|^{2}}
(q_{4}^{2}+\xi^{2}q^{2})^{3/2}} \nonumber \\
& &\times
\ln\frac{\sqrt{q_{4}^{2}+\xi^{2}q^{2}+4|\Delta^{-}|^{2}}+
\sqrt{q_{4}^{2}+\xi^{2}q^{2}}}
{\sqrt{q_{4}^{2}+\xi^{2}q^{2}+4|\Delta^{-}|^{2}}-
\sqrt{q_{4}^{2}+\xi^{2}q^{2}}} .
\label{Pi_1-gen}
\ea
From the asymptotes in Eqs.~(\ref{B16}) and (\ref{B21}), one might get the 
impression that the transversality of $\Pi_{1}^{\mu\nu}(q)\equiv 4\pi
\alpha_{s}[J^{\mu\nu}(q)+I^{\mu\nu}(q)] $ is not exact. This is just an
artifact of using the expansion for either the limit of $|q_{4}| \ll q$ or
$q \ll |q_{4}|$. To test the condition $q_{\mu}\Pi_{1}^{\mu\nu}(q)=0$, one
has to keep the subleading corrections to such expansions.

It is straightforward to check that the general expression in
Eq.~(\ref{V-4}) becomes strongly gauge dependent if the one-loop
polarization tensors $\Pi_{4}^{\mu\nu}(q)$ and $\Pi_{8}^{\mu\nu}(q)$ with
nonzero longitudinal components are used in the calculation. Moreover, the
corresponding gauge dependent contribution to the value of the mass 
comes from the region of small momenta, $|q_{4}|,q \ll |\Delta^{-}|$,
and it is logarithmically divergent,
\be
M^{2}_{\lambda} \sim \lambda^{2} \alpha_{s}^{2} \mu^{2}
\ln\frac{|\Delta^{-}|}{\epsilon_{IR}},
\ee
where $\lambda$ is the gauge fixing parameter and $\epsilon_{IR}$ is the
infrared cutoff. In deriving this result, we used the asymptotes presented 
in Eq.~(\ref{Pi_48-IR}). This result clearly demonstrates that the
presence of the longitudinal components in the polarization tensors
$\Pi_{4}^{\mu\nu}(q)$ and $\Pi_{8}^{\mu\nu}(q)$ is not acceptable. In the
complete theory these components should be zero, because the Ward
(Slavnov-Taylor) identities imply that the polarization tensor should be
transverse. As we mentioned in the main text, it is the contributions of
the (would be) NG bosons which should be added to the polarization tensors
in order to restore the transversality.

As it was already pointed out in Sec.~\ref{vac-diagram}, one has to use
the non-linear realization of color symmetry breaking for that purpose.
The polarization tensor can be found from the corresponding low energy
effective action (a similar approach for improving the polarization tensor
has been recently considered in Ref.~\cite{Gusynin:2001ib}). The effective
action for the two flavor dense QCD was described in
Ref.~\cite{Casalbuoni:2000cn}. It is given in terms of the Maurer-Cartan
one-form,
\be 
\omega_{\mu} =  {\cal V}^{\dagger} i (\partial_{\mu} 
-ig \hat{A}_{\mu}){\cal V} ,
\ee
where the field ${\cal V}$ parametrizes the coset space 
$SU(3)_{c}\times U(1)_{B}/SU(2)\times \tilde{U}(1)_{B}$.
Its explicit form reads
\be
{\cal V} = \exp\left[i \sum_{A=4}^{7}\phi^{A} T^{A}
+i\frac{\phi^{8}}{3}\left(I + \sqrt{3} T^{8} \right)\right].
\ee
In the last expression, we took into account that one of the broken
generators is $\left(I + \sqrt{3} T^{8} \right)/3$ rather than $T^{8}$.
The fields $\phi^{A}$ ($A=4, \ldots, 8$) describe dynamical (would be)
Nambu-Goldstone degrees of freedom which should necessarily appear in the
low-energy description of the color superconducting phase of dense QCD
(unless a unitary gauge is used). The interaction of gluons and would be
NG bosons with fermions is described by the following term
\cite{Casalbuoni:2000cn}:
\be
L_{f} = \bar{\psi} (i\dinot +\mu \gamma^{0} +
\gamma^{\mu} \omega_{\mu})\psi. 
\ee
Therefore, after integrating out fermions and high momenta gluons (with $q
\gg |q_{4}| \gg |\Delta^{-}|$) which are mostly responsible for generating
the gap \cite{Hong:2000fh}, the one-loop approximation of the effective
action is mimicked by the following expression:
\ba  
S_{eff} &=& -\frac{i}{2} \mbox{Tr} \ln 
\left(\begin{array}{cc} {\cal K}_{+} & \Delta\\
                  \tilde{\Delta} & {\cal K}_{-} \end{array}
\right)= -\frac{i}{2} \mbox{Tr} \ln 
\left[ {\cal K}_{-} {\cal K}_{+} \right]
-\frac{i}{4} \mbox{Tr} \ln \left[ 1
- {\cal K}_{+}^{-1} \Delta {\cal K}_{-}^{-1} \tilde{\Delta} \right] ,
\label{S-eff}
\ea
where 
\ba
{\cal K}_{+} &\equiv & i\dinot +\mu \gamma^{0}
+ {\cal V}^{\dagger} i (\dinot -ig \Ahatnot ){\cal V} , \\
{\cal K}_{-} &\equiv &i\dinot -\mu \gamma^{0}
+ {\cal V}^{T} i (\dinot +ig \Ahatnot^{T} ){\cal V}^{*} . 
\ea
The quadratic part of the induced
effective action, as follows from Eq.~(\ref{S-eff}), reads
\ba  
S^{(2)}_{eff} &\simeq& - \frac{i}{4} 
\int \frac{d^{4} q d^{4} p}{(2\pi)^{8}} \mbox{tr}
\Bigg[R_{11}\left(p\right) f_{+}(q) 
 R_{11}\left(p-q\right) f_{+}(-q)
+R_{12}\left(p\right) f_{-}(q) 
 R_{21}\left(p-q\right) f_{+}(-q) \nonumber \\
&+&R_{21}\left(p\right) f_{+}(q) 
 R_{12}\left(p-q\right) f_{-}(-q)
+R_{22}\left(p\right) f_{-}(q) 
 R_{22}\left(p-q\right) f_{-}(-q)
\Bigg], 
\label{S2-eff}
\ea
where, by derivation, 
\ba
f_{+}(q) &=& \gamma^{\mu}\left[ g A^{A}_{\mu}(q) T^{A} 
+ i q_{\mu} \sum_{A=4}^{7} \phi^{A}(q) T^{A} 
+ i q_{\mu} \frac{\phi^{8}(q)}{3} \left(I+\sqrt{3}T^{8}\right) 
\right], \\
f_{-}(q) &=& \gamma^{\mu}\left[ -g A^{A}_{\mu}(q) (T^{A})^{T} 
- i q_{\mu} \sum_{A=4}^{7} \phi^{A}(q) (T^{A})^{T} 
- i q_{\mu} \frac{\phi^{8}(q)}{3} \left(I+\sqrt{3}T^{8}\right)
\right].
\ea
(here we expanded ${\cal V}$ in powers of $\phi^{A}$).
By substituting the last expressions into Eq.~(\ref{S2-eff}), we arrive
at the result in the following form:
\ba
S^{(2)}_{eff} &\simeq& \frac{1}{2} \int \frac{d^{4} q}{(2\pi)^{4}}
\Bigg\{ \sum_{A=1}^{3} A_{\mu}^{A}(-q) \Pi_{1}^{\mu\nu}(q) A_{\nu}^{A}(q) 
\nonumber \\
&+& \sum_{A=4}^{7} 
\left[A_{\mu}^{A}(-q) -\frac{i}{g} q_{\mu} \phi^{A}(-q) \right]
\Pi_{4}^{\mu\nu}(q) 
\left[A_{\nu}^{A}(q)+\frac{i}{g} q_{\nu} \phi^{A}(q) \right]
\nonumber \\ 
&+& \left[A_{\mu}^{8}(-q) -\frac{i\sqrt{3}}{g} q_{\mu} \phi^{8}(-q) 
\right] \tilde{\Pi}_{8}^{\mu\nu}(q) \left[
A_{\nu}^{8}(q)+\frac{i\sqrt{3}}{g} q_{\nu} \phi^{8}(q) \right]
\Bigg\},
\label{S-eff-quad}
\ea
where, by derivation,
\be
\Pi_{8}^{\mu\nu}(q) \equiv \frac{1}{3} \left[
\tilde{\Pi}_{8}^{\mu\nu}(q)
+8 \pi \alpha_{s} I_{HDL}^{\mu\nu}(q) \right] .
\ee
By integrating out the $\phi^{A}$ fields, we arrive at the effective
action of the gluon field [compare with the effective action in
Ref.~\cite{Gusynin:2001ib}]:
\ba
S^{(2)}_{gl} &\simeq& \frac{1}{2} \int \frac{d^{4} q}{(2\pi)^{4}}
\Bigg\{
\sum_{A=1}^{3} A_{\mu}^{A}(-q) \Pi_{1}^{\mu\nu}(q) A_{\nu}^{A}(q)
\nonumber \\
&+& \sum_{A=4}^{7} A_{\mu\perp}^{A}(-q) \Bigg[
\Pi_{4}^{\mu\nu}(q) - \Pi_{4}^{\mu\mu^{\prime}}(q)
q_{\mu^{\prime}}\left(q_{\lambda} \Pi_{4}^{\lambda\kappa}
q_{\kappa}\right)^{-1}
q_{\nu^{\prime}} \Pi_{4}^{\nu^{\prime}\nu}(q) \Bigg]
A_{\nu\perp}^{A}(q) \nonumber \\
&+& A_{\mu\perp}^{8}(-q) \Bigg[
\tilde{\Pi}_{8}^{\mu\nu}(q) - \tilde{\Pi}_{8}^{\mu\mu^{\prime}}(q)
q_{\mu^{\prime}}\left(q_{\lambda} \tilde{\Pi}_{8}^{\lambda\kappa}
q_{\kappa}\right)^{-1}
q_{\nu^{\prime}} \tilde{\Pi}_{8}^{\nu^{\prime}\nu}(q) \Bigg]
A_{\nu\perp}^{8}(q) .
\label{S-glue}
\ea
By making use of the properties of the polarization tensor, we check that
this quadratic part of the action splits into three decoupled pieces with
different types of gluons. One of them contains the first three gluon that
do not feel the Meissner effect. The improved polarization tensors for the
five gluons corresponding to broken generators, explicitly transverse,
and thus, they can be written in the standard form:
\be
\Pi_{i,new}^{\mu\nu}(q) = 
-\Pi_{i,t}(q) O^{(1)\mu\nu}(q) 
- \frac{q_{0}^2-q^2}{q^{2}} \Pi_{i,l}(q) O^{(2)\mu\nu}(q), 
\ee
where 
\ba
\Pi_{i,t}(q) &=& -\Pi_{i}^{(1)}(q) , \\
\Pi_{i,l}(q) &=& -\frac{q^{2}}{q_{0}^2-q^2}
\left[\Pi_{i}^{(2)}(q)+\frac{[\Pi_{i}^{(4)}(q)]^2}
{\Pi_{i}^{(3)}(q)}\right], 
\ea
with $\Pi_{i}^{(x)}(q)$ ($i=4,8$ and $x=1,2,3,4$) defined in
Eq.~(\ref{Pi-i-decomp}).

In the infrared region $|q_{4}|,q \ll |\Delta^{-}|$, the improved gluon
tensors that correspond to the broken generators read 
\ba
\Pi_{4,new}^{\mu\nu}(q) 
= \frac{3}{2} \Pi_{8,new}^{\mu\nu}(q) 
= \frac{ 2 \alpha_{s} \mu^{2} }{ 3\pi } \left[
O^{(1)\mu\nu}(q)
+ \frac{q_{0}^{2}-q^{2}}{q_{0}^{2}-\frac{1}{3}q^{2}} O^{(2)\mu\nu}(q)
\right] .
\ea
Similarly, we get the expressions in other limits:
\be
\Pi_{4,new}^{\mu\nu}(q) - 4\pi\alpha_{s} I_{HDL}^{\mu\nu}(q)
=  \left\{ \begin{array}{lll}
-\frac{2 \alpha_{s} \mu^2 |\Delta^{-}|^{2}}{3 \pi q_{4}^{2}}
\ln\frac{q_{4}^{2}}{|\Delta^{-}|^{2}} \left[
O^{(1)}(q)+O^{(2)}(q) \right],
& \mbox{for} & q, |\Delta^{-}| \ll |q_{4}|; \\
\frac{\alpha_{s} \mu^2 |\Delta^{-}|^{2}}{2 |q_{4}| q}
\left[O^{(1)}(q)-O^{(2)}(q) \right],
& \mbox{for} & |\Delta^{-}| \ll |q_{4}| \ll q;\\
\frac{\pi \alpha_{s} \mu^2 |\Delta^{-}|}{q}
\left[O^{(1)}(q)-O^{(2)}(q)\right] ,
& \mbox{for} &  |q_{4}| \ll |\Delta^{-}| \ll q,
\end{array}\right.
\ee
and 
\be
\Pi_{8,new}^{\mu\nu}(q)- 4\pi\alpha_{s} I_{HDL}^{\mu\nu}(q)
=  \left\{ \begin{array}{lll}
0,
& \mbox{for} & q, |\Delta^{-}| \ll |q_{4}|; \\
\frac{2\alpha_{s} \mu^2 |\Delta^{-}|^{2}}{3 |q_{4}| q}
\ln\frac{4q_{4}^{2}}{|\Delta^{-}|^{2}}
O^{(1)}(q),
& \mbox{for} & |\Delta^{-}| \ll |q_{4}| \ll q;\\
\frac{\pi \alpha_{s} \mu^2 |\Delta^{-}|}{3q}
O^{(1)}(q) ,
& \mbox{for} &  |q_{4}| \ll |\Delta^{-}| \ll q.
\end{array}\right.
\ee
One should notice that the magnetic components [determined by the terms
with the projection operator $O^{(1)}(q)$] of the new polarization
tensors, are exactly the same as in the one-loop approximation given at
the beginning of this Appendix. The electric components are different in
general, but they appear to be also essentially the same to the leading
order approximation. Most importantly, however, the longitudinal
contributions in $\Pi_{4}^{\mu\nu}(q)$ and $\Pi_{8}^{\mu\nu}(q)$ are gone
now. The immediate consequence of this is that the vacuum energy given by
Eq.~(\ref{V-4}), as well as the masses of the pseudo-NG bosons, become
explicitly gauge invariant.

For completeness of presentation, we also note that the contribution of
the modified electric components of the polarization tensors to the
value of the pseudo-NG boson mass is of order
\be
\delta_{el} M^{2} \sim \frac{|\Delta^{-}|^{3}}{\mu}.
\ee
Because of the chemical potential in the denominator, this is clearly
a subleading correction as compared to the result in Eq.~(\ref{LOresult}).


\end{document}